\documentclass{elsart}
\usepackage{amsfonts,epsfig,float}
\newcommand{\beq}{\begin{equation}}
\newcommand{\eeq}{\end{equation}}
\newcommand{\beqa}{\begin{eqnarray}}
\newcommand{\eeqa}{\end{eqnarray}}
\newcommand{\nn}{\nonumber \\}
\newcommand {\tr}[1]{{\mathop{\mathrm{tr}}\limits_{\quad #1}} }

\def \L {\underline{\Lambda}}

\def \s {\sigma}
\def \r {\rho}

\def \Q {{\underline{\mathrm{Q}}}}

\def \t {\tau}
\def \ex {\mathrm{e}}

\def \z {\zeta}

\def \D {\Delta}

\def \la {\langle}
\def \ra {\rangle}

\def \H {{\mathcal H}}

\def \Z {{\mathbb Z}}

\def \mod {\; \mathrm{mod}\; }

\def\u1{{\widehat{u(1)}}}

\def \ch {\mathrm{ch}}

\def \Im {\mathrm{Im} \ }
\def \sgn {\mathrm{sgn} \ }
\def \MR {\mathrm{MR}}
\def \EPf {\mathrm{EPf}}
\def \Re {\mathrm{Re} \ }

\def \NUM {\mathrm{NUM}}
\begin{document}
\begin{frontmatter}
\title{The $\nu=5/2$ quantum Hall state revisited: spontaneous 
breaking of the  chiral fermion parity and  phase transition between 
abelian and non-abelian statistics}
\author[ITP,INRNE]{Lachezar S. Georgiev}
\ead{lg@thp.uni-koeln.de}
\address[ITP]{Institut f\"ur Theoretische Physik, 
Universit\"at zu K\"oln,  
Z\"ulpicher Str. 77,  \\ 50937  K\"oln, GERMANY}
\address[INRNE]{Institute for Nuclear Research and 
Nuclear Energy, Tsarigradsko Chaussee 72, \\ 1784 Sofia, BULGARIA}
\begin{abstract}
We review a recent development in the theoretical understanding of
the $\nu=5/2$ quantum Hall plateau and propose a new conformal  field
theory,  slightly different from
the Moore--Read one, to describe another universality class
relevant for this plateau.
The ground state is still given by the Pfaffian and is completely
polarized, however, the elementary
quasiholes are charge $1/2$ anyons with abelian statistics
$\theta=\pi/2$, which obey complete spin--charge separation.
The physical hole is represented by two such
quasiholes plus a free neutral Majorana fermion. 
We also compute the periods and amplitudes of the chiral 
persistent currents in both states and 
show that they have different temperature 
dependence. 
Finally, we find indications of a classical two-step phase transition
between the new and the  Moore--Read states, through a
compressible state, which is characterized by the
spontaneous breaking of a hidden $\Z_2$ symmetry corresponding to 
the conservation of the chiral fermion parity.
We believe that this transition could explain the ``kink" observed
in the activation experiment for  $\nu=5/2$.
\end{abstract}
\end{frontmatter}
\section{Introduction}
The nature of the first observed fractional quantum Hall (FQH)  state,
with even denominator of the  filling factor $\nu=5/2$ \cite{willett,pan},
remains obscure more than a decade.
Its collapse with the increase of the tilted field
\cite{eisen88}
suggested that its ground state might be a spin-singlet, however,
this turned out to be wrong.
Shortly after the first challenge of  this interpretation
\cite{morf} a new experiment with the $\nu=5/2$ FQH state
\cite{pan2} confirmed that its ground state is indeed
spin-polarized.
The first spin-singlet state for the $\nu=5/2$ plateau,
motivated by the wrong interpretation of the tilted field
experiment, was introduced by Haldane and Rezayi  \cite{hr} 
and besides other inconsistencies,
such as the violation of the spin--statistics relation, non-unitarity and
absence of modular invariance, 
 turned out to be an excited state over the 331 ground
 state \cite{cgt}; later it was found to describe a
compressible  state \cite{read-green,read_5-2} at the transition between
weak and strong pairing phases of the p-wave BCS Hamiltonian,
explaining why the so called  Haldane--Rezayi state \cite{hr} 
is not a true FQH state.

The 1+1 dimensional conformal field theory (CFT), describing
the FQH edge excitations in the thermodynamic limit,
has proven to be a convenient tool for analyzing the
universality classes of the FQH states
\cite{fro2000,cgt,cgt2000}.
The CFT approach became even more important after the recent experiment
\cite{grayson} showing that the chiral Luttinger liquid point of view 
is  at best incomplete or at worst wrong.
So far the most successful CFT for the $\nu=5/2$ FQH state
has been proposed  by Moore--Read (MR) \cite{mr,read_5-2}.
Numerical calculations \cite{5-2hr} have shown that, 
after proper particle--hole (PH) symmetrization,
the MR state seems to have a good overlap with the exact ground state
for the $\nu=5/2$ plateau, 
and that this state most likely determines the universality class
of the $\nu=5/2$ FQH state  for zero temperature. However, since the 
PH symmetry
is a characteristics of the universality class,  it is rather strange
that the MR Hamiltonian used in \cite{5-2hr} is not PH symmetric. 
In this paper we try to interpret
the results of \cite{5-2hr} in a way to  explain this peculiarity.

Another striking fact become apparent after the recent activation
experiment \cite{pan}. The logarithmic plot of the longitudinal 
resistance $R_{xx}$, as a function of the inverse temperature, 
turned out to be non-linear. The slope suddenly changes 
around $T\sim 15$ mK \cite{pan}, which was called ``a kink" that  
these authors could not explain. We stress that the change of
 the slope of the 
diagonal resistance is a clear indication of a phase
transition implying that there should be another incompressible phase 
at $\nu=5/2$, with a different energy gap, which like the MR state, 
is supposed to be completely polarized \cite{pan2}. 
In other words, there should be another universality
class, hence another (rational) CFT, which is relevant for the 
$\nu=5/2$ plateau. 
In this paper we propose a new universality class for the
$\nu=5/2$ FQH plateau ---  a rational extension of the MR 
state --- which we call the Extended Pfaffian (EPf) state.
Although it may look innocent, this extension changes dramatically 
the structure of the excitations.
In Sect.~\ref{sec:epf} we analyze the  CFT for the EPf state and 
show that  the minimal electric charge in this state  is $1/2$
(in units in which the electron charge is $-1$)
and the quasiparticle's statistics is abelian.
This has to be compared with the MR state where the minimal electric 
charge  is  $1/4$ and the statistics of quasiparticles is 
non-abelian \cite{mr,fra-nay-schout,fra-huer-zem,maiella}.
We recall that, the construction of the  EPf state, together 
with the assumption that the energy gap in the
FQH fluids  has a universal component proportional to the quasihole's
electric charge, was able to explain \cite{gaps} the non-monotonic
structure \cite{pan} of the parafermionic Hall states in the
second Landau level.

One more peculiarity of the MR state is the fact that 
\textit{its chiral fermion parity number is not conserved in the twisted 
sector}, due to the non-abelian fusion rules of the quasiparticles,
hence, the fermion parity is spontaneously broken in this sector. 
In Sect.~\ref{sec:parity} we show that this reduces
the topological order of the MR state  as compared to the 
double-layer  331 state \cite{milo-read,cgt}.
Recall that the MR state was interpreted \cite{cgt,read-green,cabra} 
as a low-barrier (or high-tunneling) limit of the 331 state and 
the transition from 331 to the MR state could be 
characterized by the spontaneous breaking of this $\Z_2$ symmetry.
Note that the fermion parity plays a fundamental role in finite 
geometries since  the ground states of interacting fermionic 
systems are believed to be paramagnetic/diamagnetic for even/odd 
fermion parity
\cite{loss-parity,loss-goldbart}, in very much the same way
like the free systems \cite{byers-yang};  the latter is 
known as the \textit{Leggett conjecture} \cite{leggett}. 

Our analysis shows that the kink observed  in the activation 
experiment at $\nu=5/2$  \cite{pan} is due to a classical two-step 
phase transition
involving an intermediate compressible state. At low temperatures
the system is in the MR phase, in which the chiral fermion parity 
$\Z_2$ symmetry  is spontaneously broken. When the temperature becomes
bigger than the activation energy for  the MR state, the system undergoes
 a second order phase transition (in which the $\Z_2$ symmetry is 
restored) to a compressible state, which is topologically equivalent to
the Composite Fermions (CF) Fermi liquid found before
\cite{morf,5-2hr,fermi5-2}, and then a first order phase transition to 
the EPf state. 
Using the previous estimation of the energy gaps according to the
gap ansatz in \cite{gaps} we find in Sect.~\ref{sec:gaps} that the 
gap of the EPf state
is almost 3 times bigger than that of the MR state, i.e., 
$\D_\EPf=110$~mK and $\D_\MR=33$~mK for the sample of \cite{pan}, 
which could explain the kink observed in  the activation 
experiment \cite{pan}, 
as well as the absence of fermion parity number in the MR state.
It is also intriguing that similar  
kinks are seen also in the activation experiments for the neighboring
FQH plateaux at $\nu=7/3$ and $8/3$, which suggests that this is 
probably a general phenomenon.

In Sect.~\ref{sec:PH} we discuss the 
PH symmetry of the MR and EPf states and reinterpret the numerical results
of \cite{5-2hr} in order to explain the absence of the PH symmetry 
in the MR state. In Sect.~\ref{sec:heat} we compute the contributions of 
all topologically inequivalent quasiparticles to the specific heat of the
EPf state.

In Sect.~\ref{sec:pers} we describe how to derive the mesoscopic 
persistent currents of MR and EPf states directly from their effective 
CFTs, using the modular covariance of the CFT characters, in which
the non-analytic factors of Cappelli--Zemba (CZ) \cite{cz} are 
included. We stress again
that this approach is a complement and an alternative to 
the chiral Luttinger liquid description which is 
unsatisfactory \cite{grayson}.
We find that the persistent currents in the MR and EPf states are 
periodic functions of the magnetic flux with period exactly one
flux quantum. The amplitudes of the persistent currents are the same at 
zero temperature, however, \textit{for finite temperatures the 
amplitude of the persistent current in the EPf state is always bigger
 than that in the MR state}. 
This fact could be used in principle to detect any transition between 
the two states, e.g., in a SQUID experiment \cite{pers-exp}. Both currents
exhibit a universal non-Fermi liquid temperature behaviour and we 
find analytic formulae for the temperature dependence of the amplitudes 
in both limits of low and high temperatures.

In Sect.~\ref{sec:PT} we describe in more detail the above mentioned 
scenario for the phase transition between the MR and EPf states. 
Again we point out that this transition could possibly explain the 
kink observed in the activation experiment \cite{pan}. 
Some technical remarks are summarized in several appendices.
\section{The Moore--Read state: the CFT coset point of view}
\label{sec:mr}
Originally, the MR ground and excited states  were
 introduced \cite{mr}  as correlators
of certain operators in a CFT based on the chiral algebra
$\u1\otimes \mathrm{Ising}$. For example, the ground state wave function
 (up to the standard Gaussian exponent) is expressed 
as a correlator of $2N$ ``electron operators"
\cite{mr,milo-read,cgt}
\begin{equation}\label{el}
\psi_{\mathrm{el}}(z)= \varphi(z) :\ex^{-i\sqrt{2}\phi(z)}:,
\end{equation}
where $\varphi(z)$ is the  neutral Majorana fermion in the Ising model and
$:\ex^{-i\sqrt{2}\phi(z)}:$ is the $\u1$ vertex exponent representing
the charged component of the electron \cite{cgt}, as follows
\begin{equation}\label{pfaff}
\Psi_{\mathrm{GS}}(z_1,\ldots, z_{2N})=
\la \prod_{j=1}^{2N} \psi_{\mathrm{el}}(z_j)  \ra =
\mathrm{Pf}\left( \frac{1}{z_i-z_j}\right) \prod_{i < j}
(z_i-z_j)^2 .
\end{equation}
This polarized state was called paired
in analogy with the real space BCS type wave function expressed
by the Pfaffian \cite{mr,cgt}.
The stress energy tensor of the MR state is a sum of
the $\u1$ and the Ising contributions
\beq\label{T}
        T(z)=-\frac{1}{2}:\partial \phi(z)^2:
        - \frac{1}{2} :\varphi(z)
        \partial \varphi(z):
\eeq
and has a central charge $c_\MR=3/2$.
The elementary quasiholes are represented by
\begin{equation}\label{MR-q.h.}
\psi^\MR_{\mathrm{q.h.}}(z)= \s(z)
:\ex^{i\frac{1}{2\sqrt{2}}\phi(z)}:,
\quad Q^\MR_{\mathrm{q.h.}}=1/4, \quad \D^\MR_{\mathrm{q.h.}}=1/8,
\end{equation}
where $\s$ is the chiral spin--field of the Ising model with 
(neutral) CFT dimension $1/16$ 
and the normal ordered exponent represents the
charged component of the quasihole.
The quasiholes (\ref{MR-q.h.}) have  rather peculiar properties ---
their electric charge $Q^\MR_{\mathrm{q.h.}}$  differs from the
denominator of the filling factor, they carry half-integer  flux
$\Phi^\MR_{\mathrm{q.h.}}=1/2$,  so that must be created in
pairs, their total CFT dimension is $\D^\MR_{\mathrm{q.h.}}=1/8$, however
they obey  non-abelian statistics \cite{mr,schout2,cgt,cgt2000}.
Note that for QH states with  even-denominator $d_H$  the minimal 
electric charge  is $Q_{\min}=1/(l d_h)$, where  $l$ is an  even integer 
called  the \textit{charge parameter}  ---  see Theorem 4.3 
in \cite{fro-stu-thi}.
A recent study of the MR state in the framework
of the parafermionic FQH states \cite{rr,fra-huer-zem,maiella},
realized as coset constructions \cite{cgt2000}, has  made
the mechanism of clustering more transparent. 
In particular, the MR state is
realized as an affine coset  projection \cite{cgt2000} 
$\widehat{su(2)}_1\oplus\widehat{su(2)}_1  \to
\left(\widehat{su(2)}_1\oplus\widehat{su(2)}_1 \right)/ \widehat{su(2)}_2$,
in  an abelian lattice
theory of the type \cite{cgt2000,fro-stu-thi}
$(3| {}^1A_1 {}^1A_1)$,  i.e., a  theory with a $K$-matrix and
a charge vector $\Q$
\[
        K=\left[\matrix{3 & 1 & 1 \cr 1 & 2 & 0 \cr 1 & 0 & 2}\right],
        \quad \Q=(1,0,0),
\]
which is interpreted as removing the layer (or color) $su(2)$ symmetry.
The above pair $(K,\Q)$ uniquely determines the Chern--Simons topological
theory in the bulk and the coset projection is interpreted as gauging
out the  $\widehat{su(2)}_2$ layer symmetry, which produces a non-abelian
topological effective theory for the bulk of the FQH fluid.
The Chern--Simons effective field theory describing the bulk of 
the MR state has attracted much attention 
\cite{fra-huer-zem,fra-nay-schout} due to the peculiar properties 
 of the non-abelian quasiparticles present there.
We stress that the CFT on the $1+1$ dimensional boundary determines
uniquely the topological field theory in the bulk \cite{fro2000}.
Therefore,  the pairing rule in the MR state 
(see Eq.~(\ref{pr}) below), which gives rise to a $\Z_2$ orbifold CFT
construction \cite{milo-read,cgt},  has crucial implications for the bulk 
effective field theory of the fermionic Pfaffian state.

The quasiholes wave functions, after the coset projection, 
can be obtained by symmetrization \cite{cgt2000} of the wave 
functions of their parent counterparts and
the non-abelian statistics is understood as a result of degeneration
in the space of quasihole wave functions during this process.
Another advantage of this projective construction is that the projected 
model inherits some structures, such as pairing rules and modular 
covariance,  from its parent \cite{cgt2000}.

The chiral  partition functions,
$\chi_{l,\rho}(\t,\z)= \tr{\H_{l,\rho}} 
\left( q^{L_0-c/24}\ex^{2\pi i \z J_\mathrm{el})} \right)$, 
for a disk sample with $L_0$ and $J_\mathrm{el}=(\Q|\underline{J}_0)$ 
being the zero modes of the
stress tensor (\ref{T}) and the electric current, respectively
\cite{cz,cgt,fro-stu-thi},
representing all topologically inequivalent quasiparticles
\cite{cgt2000,gaps}, can be
obtained  from Eq. (10) in \cite{gaps}
(for $k=2$, $M=1$ and denoting by
        $\ch_0:=\ch(\L_0+\L_0)$, $\ch_{1/16}:=\ch(\L_0+\L_1)$ and
        $\ch_{1/2}:=\ch(\L_1+\L_1)$ the Ising model characters with
        lowest CFT dimensions $0$, $1/16$ and $1/2$, respectively)
\begin{eqnarray}
\chi_{2l,0}(\tau,\zeta) &=& \ex^{-\frac{\pi}{2} \frac{(\Im\z)^2}{\Im\t}}
 \left( K_{2l}(\tau,2\zeta;8)\ch_{0}(\tau)+K_{2l+4}(\tau,2\zeta;8) 
\ch_{1/2}(\t) \right) ,\nn
 && \mathrm{where} \ \ l=-1,0,1,2,   \nn
\chi_{\pm 1,0}(\tau,\zeta)&=& \ex^{-\frac{\pi}{2}\frac{(\Im\z)^2}{\Im\t}}
\Bigl( K_{\pm 1}(\tau,2\zeta;8)+K_{\mp3}(\tau,2\zeta;8) \Bigr)
\ch_{1/16}(\tau) \label{chi_MR} = \nn
&=& \ex^{-\frac{\pi}{2}\frac{(\Im\z)^2}{\Im\t}}
K_{\pm 1/2}(\tau,\zeta;2) \ch_{1/16}(\tau),
\end{eqnarray}
where the $K$-functions are the $\u1$ chiral partition functions
\cite{cgt,cgt2000,gaps}
\begin{eqnarray}\label{K_l}
K_l(\tau,\zeta;m)&=&\frac{1}{\eta(\tau)}\sum_{n\in \Z}
q^{\frac{m}{2}(n+\frac{l}{m})^2}
\mathrm{e}^{2\pi i \zeta (n+\frac{l}{m})}, \nn
\eta(\tau)&=&q^{\frac{1}{24}} \prod_{n=1}^\infty (1-q^n),\quad
q=\ex^{2\pi i \tau}
\end{eqnarray}
and the Ising model characters are given by \cite{cgt,cgt2000}
\beqa\label{Ising}
\ch_{0}(\tau) &=& \frac{q^{-\frac{1}{48}}}{2}
        \left( \prod_{n=1}^\infty (1+q^{n-\frac{1}{2}}) +
        \prod_{n=1}^\infty (1-q^{n-\frac{1}{2}}) \right), \nn
\ch_{1/2}(\tau) &=& \frac{q^{-\frac{1}{48}}}{2}
        \left( \prod_{n=1}^\infty (1+q^{n-\frac{1}{2}}) -
        \prod_{n=1}^\infty (1-q^{n-\frac{1}{2}}) \right), \nn
 \ch_{1/16}(\tau) &=& q^{\frac{1}{24}} \prod_{n=1}^\infty (1+q^n).
\eeqa
The modular parameters $\tau$, $\z$
are related to the inverse temperature $\beta=1/k_B T$ and
the magnetic flux $\Phi$ (cf. \cite{cz}) as follows
$2\pi R /v_F \, \mathrm{Im} \tau = \beta$,
$2\pi \mathrm{Im} \, \zeta=\beta \Phi$ ($R$ is the radius of the edge
and $v_F$ the Fermi velocity).

The  partition function for an annulus sample can be written
as a bilinear combination  \cite{cz} of the characters (\ref{chi_MR})
\begin{equation}\label{Z_MR}
Z_{\mathrm{MR}}(\tau,\zeta)=
\sum_{l=0,\pm 1, \pm 2, 4} \vert \chi_{l,0}(\tau,\zeta)\vert^2
\end{equation}
and is invariant \cite{cgt}  with respect to the modular 
transformations \cite{cz}
$T^2: \tau\to \tau+2$, $S: \tau\to -1/\tau$, $U:\zeta\to \zeta+1$
and
\begin{equation}\label{V}
V: \qquad \zeta \to \zeta +\tau  \quad
\Longleftrightarrow \quad  N_{\Phi} \to N_{\Phi}+1  .
\end{equation}
The $V$ transformation, Eq.~(\ref{V}), represents increasing the 
flux through the sample
$N_\Phi=\Phi/ \phi_0$ by one unit $\phi_0=h/e$ and therefore the
Hall current is realized as the Laughlin spectral flow \cite{cz}
under $V$.
Note that the $V$-invariance of the partition function (\ref{Z_MR})
requires multiplying the characters (\ref{chi_MR}) by the non-analytic
factor of Cappelli--Zemba  \cite{cz}
\beq\label{CZ}
\exp\left( -\pi \ \nu \ \Im \t \ 
\left( \frac{\Im\z}{\Im\t} \right)^2 \right)
\eeq
and we have to stress that \textit{without the factor (\ref{CZ}) in the 
CFT approach it is not possible to reproduce the persistent 
currents correctly}, see Sect.~\ref{sec:pers}.
 
The filling factor is derived as the electric charge transferred
between the edges under the spectral flow (\ref{V}).
The topological order of the MR state, i.e., the number of 
topologically inequivalent charged excitations with an absolute value of 
the electric charge less than one \cite{fro2000,cgt}, is 
$\mathrm{TO}_\MR=6$ according to Eq. (\ref{Z_MR}).

The fusion rules, i.e., the rules for making composite
quasiparticles are complex due to the non-abelian fusion
rules in the Ising model and can be found in
\cite{cgt,cgt2000}.
Here we point out one peculiar property of the MR state --- the
multi-electron wave functions, in the Ramond sector
(with partition function  $\chi_{1,0}$) of the Ising model,
do not have a definite chiral fermion parity\footnote{I thank 
Ivan Todorov for pointing out this fact to me} as a result of the
non-abelian fusion rules of the spin fields $\s$ 
\cite{milo-read,cgt,cgt2000}
\beq\label{fusion}
\s \times \s \simeq 1+\varphi,
\eeq
which mixes states with opposite fermion parities.  
This leads to the spontaneous breaking of the chiral fermionic parity
in the MR state, which is investigated in Sect.~\ref{sec:parity}.
\section{The Extended Pfaffian state}
\label{sec:epf}
In this section we shall illustrate the general procedure of
local chiral algebra  extension \cite{gaps} on the example of the 
$k=2$ Read--Rezayi state, i.e., the extension of the MR state.
According to \cite{cgt2000}, the Ising factor of the MR chiral algebra
of the  $k=2$ parafermion FQH state is realized as a diagonal
affine coset
\beq\label{ising}
\left( \u1 \otimes \mathrm{Ising}\right)^{\Z_2},
\quad
\mathrm{Ising}=
\left(\widehat{su(2)}_1\oplus\widehat{su(2)}_1\right) / \widehat{su(2)}_2,
\eeq
and the quasiparticle excitations of the MR state
can be labelled by the
$\u1$ charge $l=-3,\ldots,4$
and the Ising model field $\Phi_{\mathrm{I}} \in \{1,\s,\varphi \}$.
The $\Z_2$ superscript in Eq. (\ref{ising}) expresses a $\Z_2$
selection rule
(called a parity rule in \cite{cgt,cgt2000}), which states that
the tensor product
$:\ex^{i \frac{l}{2\sqrt{2}}\phi(z)}: \otimes \, \Phi_{\mathrm{I}}(z)$
of  $\u1$ and Ising models excitations  with the label
$(l,\Phi_{\mathrm{I}})$ is an admissible excitation
 of the FQH  fluid iff
\beq\label{pr}
P[\Phi_{\mathrm{I}}]=l \mod 2, \quad
\mathrm{where} \quad  P[1]=P[\varphi]=0,\;\; P[\s]=1   .
\eeq
The $\Z_2$ number $P$ in Eq. (\ref{pr}) is defined  according to
\cite{cgt2000,gaps} as
$P[\L_\mu+\L_\r] = \mu + \r \mod k$, for $k=2$, taking into account
the identification \cite{cgt2000,gaps}
$\Phi_{\mathrm{I}}(\L_0+\L_0)=1$,
$\Phi_{\mathrm{I}}(\L_0+\L_1)=\s$, $\Phi_{\mathrm{I}}(\L_1+\L_1)=\varphi$.
The gluing condition (\ref{pr}) is the price one has to pay for
splitting the
neutral and charged degrees of freedom \cite{cgt,cgt2000}
and expresses the absence
of complete spin--charge separation in the MR state.
However, Eq. (\ref{pr}) means that the Majorana fermion $\varphi$,
which has
$P[\varphi]=0$, can be glued to the $l=0$ $\u1$ exponent, i.e., to
the identity. Therefore, the Majorana fermion exists as a free
 neutral excitation on the edge of the MR state.
Moreover, it carries no flux and need not to be paired \cite{gaps}.
Note that $\varphi$ has a CFT weight $1/2$ and its correlation functions
are single-valued.
 Therefore, \textit{we claim \cite{gaps} that it should be
added to the chiral observable algebra, leading to a
$\Z_2$ extension of the MR chiral algebra, which
produces the EPf state}. We stress that the EPf state is expected to be
more stable than the MR \cite{gaps,fro2000} due to the extension
of the chiral algebra. Note that the Majorana fermion $\varphi$ 
could be interpreted as the result of the fusion of one 
electron (\ref{el}) and a $2$ flux quanta composite 
$:\ex^{i\sqrt{2} \phi(z)}:$, both being legitimate excitations of the QH 
system.

The electron operator in the EPf state is
the same like that in the MR state, given by Eq.~(\ref{el}),
and the ground state of the EPf model
is still given by Eq. (\ref{pfaff}).
Also, since the CFT dimension of the electron is still  $3/2$,
like in the MR state, the tunneling current--voltage relation remains the
same, i.e., $I\sim V^3$ (neglecting the lowest Landau level
contribution \cite{read_5-2}).
In addition to the neutral Majorana fermion, in the EPf state,
there are anyons $:\ex^{\pm i \sqrt{2}\phi(z)}:$,
corresponding to $l= \pm 4$
in Eq. (\ref{pr}),  also freely available at the edge since
they can be glued to  the identity in the Ising model.
The stress energy tensor in the EPf state is the same like in the MR
state, Eq.~(\ref{T}), since we only added the neutral Majorana fermion
to the MR chiral algebra, which already contains this
stress tensor   and its central charge is $c_{\EPf}=3/2$.

However, the charged excitations in the EPf state have a different
structure.
After the extension, all excitations should  also be local with respect
to the neutral Majorana fermion $\varphi$ \cite{gaps} and this leads to the
so called ``even-charge" restriction \cite{gaps}. In the $k=2$ case this 
simply means that the spin-field $\s$ is no longer a legal excitation of
the extended chiral algebra since it is not relatively local with respect
to the Majorana fermion, i.e., their wave functions are not single
 valued in the coordinates of $\varphi$.
In other words, we should  treat $\varphi$ on the same footing as the
physical electron. Therefore, the lowest-charge admissible
 excitation of the EPf state appears to be
not the MR quasihole (\ref{MR-q.h.}) but
\begin{equation}\label{EPf-q.h.}
\psi^\EPf_{\mathrm{q.h.}}(z)= :\ex^{i\frac{1}{\sqrt{2}}\phi(z)}:,
\quad Q^\EPf_{\mathrm{q.h.}}=1/2, \quad \D^\EPf_{\mathrm{q.h.}}=1/4
\end{equation}
carrying electric charge $1/2$ , magnetic flux  
$\Phi^\EPf_{\mathrm{q.h.}}=1$ and having  a (total) CFT dimension
$\D^\EPf_{\mathrm{q.h.}}=1/4$.
Recall that for general $k$-even the above exponent should be glued to
$\Phi(\L_0+\L_2)$ \cite{gaps}, however, for $k=2$ the latter field
coincides with the identity.
Note also that Eq. (\ref{EPf-q.h.}) corresponds to the choice $l=2$ in
(\ref{pr}) and  is another legitimate excitation of the EPf fluid.
This means that \textit{all excitations of the EPf state, unlike those of the
MR state,   satisfy a (chiral)
spin--charge separation}, which  is one of the well-known patterns of
quantum numbers fractionalization in the FQH effect.
This is natural since the electron itself splits into a ``holon"
 and a ``spinon", which then move independently on the edge.
The quantum statistics of the EPf quasihole  $\psi^\EPf_{\mathrm{q.h.}}(z)$
 is abelian, according to Eq. (\ref{EPf-q.h.}), and is equal to
(twice) its CFT dimension, i.e.,  $\theta=\pi/2$.
The electron (\ref{el}), in the EPf fluid,
decays into 2 quasiparticles plus
one Majorana fermion. Indeed, the fusion of two quasiparticles
$\ex^{-i\frac{1}{\sqrt{2}}\phi(z)}\
\ex^{-i \frac{1}{\sqrt{2}}\phi(w)} \simeq (z-w)^{-1/2}
\ex^{-i \sqrt{2}\phi(w)} $ reproduces the charged part of the
electron, which is a boson, while the fermionic statistics is recovered
by the Majorana fermion.

The chiral partition functions for the EPf state are expressed as sums
of those of the MR state, see Eq. (23) in \cite{gaps}, and the non-chiral
partition function is their bilinear combination
\beqa
\widetilde{\chi}_{2l,0}(\t,\z)&=&
\ex^{-\frac{\pi}{2} \frac{(\mathrm{Im} \ \z)^2}{\mathrm{Im} \ \t}}
K_l(\t,\z;2)\left( \ch_0(\t)+\ch_{1/2}(\t)\right) ,\label{chi_EPf}
\\ 
Z_\EPf&=&\sum_{l=0}^1 \vert\widetilde{\chi}_{2l,0}(\t,\z)\vert^2.
\label{Z_EPf}
\eeqa
We stress that the partition function (\ref{Z_EPf}) is 
a \textit{weak modular invariant} \cite{cz}, i.e., it is invariant
under $T^2$, $S$ ,$U$ and $V$ transformations (\ref{V}) 
(see Appendix~\ref{app:mod}),
which is one of the necessary conditions for the EPf CFT to describe  
an acceptable FQH state \cite{cz,fro2000}.
In particular, due to the Verlinde fusion rules formula \cite{CFT-book}, 
the $S$-invariance guarantees that the spectrum
of quasiparticle is closed, i.e., when quasiparticles are fused together
 or when temperature increases no other topological excitations should 
appear than these already described. 
In addition, the unitarity of the $S$-matrix ensures that
the excitations' spectrum is complete since the quantum dimensions
$D_i$ \cite{CFT-book} of the quasiparticles should satisfy 
\[
\sum_i (D_i)^2=\frac{1}{S_{00}^2}, \quad \mathrm{where} 
\quad D_i=\frac{S_{0i}}{S_{00}}>0,
\] 
$S_{ij}$ is the $S$-matrix acting over the characters and $0$ 
corresponds to the vacuum character. Note that $S_{00}$ is determined 
entirely in terms of the chiral algebra, i.e., independently of the 
set of the excitations.

We stress that the chiral fermion parity, which can be defined here as 
$P/2 \mod 2$,  is conserved in the EPf state,
as explained in \cite{gaps}, in contrast to the MR state.
Moreover, it seems that the conservation of  chiral fermion parity in the
MR state is not consistent with the modular invariance, see 
Sect.~\ref{sec:parity}.

All fusion rules for the EPf quasiparticles  are abelian
(recall that the Majorana fermion itself satisfies abelian fusion rules
$\varphi \times \varphi =1$).
 Note that the abelian statistics of the quasiparticles
does not contradict to the half-integer central charge of the
Virasoro algebra.
The basic quasiparticles in both the EPf and the MR states
satisfy the generalized charge--statistics relation found in
\cite{gaps}.
 The topological order of the EPf state is  $\mathrm{TO}_\EPf =2$
 after the extension, as seen from (\ref{Z_EPf}).
According to the stability criteria S1--S3 in \cite{fro2000},
the EPf state is expected to be more stable at higher temperature 
than the MR state,
resp. to have a bigger energy gap, due to the smaller topological
order. Finally, we note that the new RCFT described in this section 
defines a new universality class relevant for the description of
the FQH state at $\nu=5/2$. 
As far as I know, this is the first proposal for another polarized 
state at this filling factor, that is different from the MR state, 
the existence of which is suggested by the activation 
experiment \cite{pan}. 
\section{Energy gaps for the MR and EPf states}
\label{sec:gaps}
In what follows we shall need some estimates of  the energy gaps 
for the MR and the EPf states. To this end we 
use our previous analysis \cite{gaps} of the energy gaps in 
the parafermionic hierarchy, which are based on the stability
criteria in S1--S3 in \cite{fro2000}. In particular,
the measurable energy gap of the $\nu=5/2$ state for the sample
of \cite{pan},  according to Eq.~(5) in \cite{gaps}, is given by
\beq\label{gap}
\D=\alpha \frac{e^2}{4\pi\epsilon l_B} 
\Delta_{\mathrm{q.h.}} -\Gamma,
\quad \alpha=0.0063, \quad \Gamma=0.045\ \mathrm{K}
\eeq
We point out that the gap estimated in \cite{gaps} for the $\nu=5/2$ 
plateaux should correspond to the EPf state since the minimal electric 
charge after the extension\footnote{this extension is necessary to explain
the non-monotonic structure of the parafermionic hierarchy \cite{gaps}} is 
$Q_{\mathrm{q.h.}}^\EPf=1/2$, and the CFT dimension of the quasihole is  
$\Delta_{\mathrm{q.h.}}^\EPf=1/4$, see Table~1 in \cite{gaps}, i.e.,
\beq\label{gap_EPf}
\D_\EPf=\alpha \frac{e^2}{4\pi\epsilon l_B} \frac{1}{4} 
-\Gamma = 110\ \mathrm{mK}.
\eeq
Recall that the value $110$ mK in Eq.~(\ref{gap_EPf}) is not a 
prediction at all  since it was used, together with the experimental
gap measured for $\nu=8/3$, in \cite{gaps} to fit the parameters
$\alpha$ and $\Gamma$ in Eq.~(\ref{gap}).
Next, according to Eqs.~(\ref{EPf-q.h.}) and (\ref{MR-q.h.}) 
the quasihole's 
CFT dimension for the MR state  is half that for the EPf state, i.e., 
$\Delta_{\mathrm{q.h.}}^\MR=\Delta_{\mathrm{q.h.}}^\EPf /2$  
and since the magnetic length $l_B$ is the same we can write  
\beq\label{gap_MR}
\D_\MR=(\D_\EPf+\Gamma)/2-\Gamma\simeq 33\ \mathrm{mK}
\quad (\pm 10 \ \mathrm{mK} )
\eeq 
for the sample of \cite{pan}. Note that Eq.~(\ref{gap_MR}) is a 
true prediction that can be used a s test for the assumptions made 
 in \cite{gaps}, see Sect.~\ref{sec:PT}.
To the best of my knowledge, this is the first analytic estimate of the 
second energy gap for the $\nu=5/2$ FQH state in the sample of \cite{pan}.
We recall that the pure energy gap (\ref{gap}), i.e., the gap for the 
system without disorder, $\Gamma=0$, is proportional to the CFT 
dimension of the quasihole only in first approximation \cite{gaps}. In general we should expect up to $30 \%$ deviation \cite{jain-goldman}, 
which justifies the appearance of $\pm 10$~mK in Eq.~(\ref{gap_MR}). 
Nevertheless, 
this precision seems to be sufficient to determine the relative 
structure of the incompressible states at $\nu=5/2$.
The important issue is that the energy gap for the EPf state is 
significantly bigger than that of the MR state and the consequences of 
this CFT-based conclusion are addressed in Sect.~\ref{sec:PT}.
\section{Spontaneous breaking of the chiral fermion parity in the 
MR state}
\label{sec:parity}
The effective CFT Hamiltonian of the chiral QH fluid corresponding 
to the MR state on a disc with radius $R$
\beq\label{ham}
H_\mathrm{CFT}= \frac{v_F}{R} \oint \frac{d \, z}{2\pi i}\  z \ T(z),
\quad z=\exp\left(\frac{v_F\, t-i\, x}{R}\right),
\eeq
where  $T(z)$ is given by Eq.~(\ref{T}) and $v_F$ is the edge Fermi 
velocity, is invariant with respect
to the transformation $\gamma_F=\gamma_F^\dagger = \gamma_F^{-1}$,
which changes  the sign of the fermion  fields
\beq\label{g_f}
\gamma_F \ \varphi(z) \  \gamma_F^{-1} = - \varphi(z) \quad \Longrightarrow
\quad \gamma_F \ H_\mathrm{CFT} \ \gamma_F^{-1} = H_\mathrm{CFT}.
\eeq
Since $\gamma_F^2=1$, Eq.~(\ref{g_f}) defines a $\Z_2$ group of 
symmetry  transformations of the MR Hamiltonian. 
Therefore the eigenstates of the Hamiltonian (\ref{ham}) 
belong to representation spaces of $\Z_2$ and 
the CFT characters can be assigned a ``good" quantum number, 
the chiral fermion parity,
which is supposed to be preserved by the dynamics.
However, as we have shown in Sect.~\ref{sec:mr}, such a quantum
number cannot be preserved by the MR fusion rules,
since the ground state $\s(0)|0\ra$ has no definite fermion parity 
number 
due to the fact that $\s$ cannot be assigned any  $\gamma_F$
number because of the non-abelian fusion rules (\ref{fusion}). 
Therefore the above $\Z_2$ symmetry is spontaneously broken
in the MR state.

We stress that, according to the discussion at the end of 
Sect.~\ref{sec:epf}, the $\Z_2$ symmetry (\ref{g_f})  is 
not the usual symmetry of the Ising model, the quantum numbers of which 
are given in Eq.~(\ref{pr}).

This spontaneous breaking  could be revealed by comparing 
the quantum numbers of the topological excitations of the MR state
to those of the double-layer 331  FQH state.
To this end we recall that according to \cite{cgt} 
(see also \cite{read-green,cabra}) the MR state could be interpreted 
as a high-tunneling limit of the  
 331 state. The number $4$ in the topological degeneracy $4m$
on the torus  of the 331 state (for which $m=2$)
could be interpreted as  coming from a 
$\Z_2\times \Z_2$ symmetry,
whose quantum numbers are implicit
 in the character formulae 
(see Eq.~(2.14) in \cite{cgt})
\beq\label{331}
\ch_{\lambda}^{331}(\t,\z)=K_{\lambda}(\t,0;4)K_{\lambda}(\t,2\z;8)+
K_{\lambda+2}(\t,0;4)K_{\lambda+4}(\t,2\z;8).
\eeq
The first $\Z_2$ quantum number characterizes the boundary 
conditions, which are untwisted for even $\lambda$ and twisted
for odd $\lambda$, while the second one is  the $\Z_2$ parity
taking values even/odd \cite{milo-read}. 
It is worth mentioning that the 331 state has a  topological 
structure slightly different from that of the MR state \cite{cgt}
since its annulus partition function has two more terms,
denoted by $\lambda=\pm 3$ in \cite{cgt}, which would have corresponded 
to $l=\pm 3$  in Eq.~(\ref{Z_MR}) for  the MR state. Although the
corresponding two 331 characters are formally the same like the 
$\lambda=\mp 1$ ones,
they are  distinguished by their fermion parity.
Indeed, the Dirac--Weyl characters 
$K_{\pm 1}(\t,0;4)$ including the fermionic zero 
modes, which appear in the twisted characters Eq.~(\ref{331}) 
for $\lambda=\pm 1, \pm 3$, could be assigned fermionic numbers 
$F=\pm 1/2$, i.e., $\gamma_F=(-1)^{\pm 1/2}$, 
in agreement with the general discussion in  Sect.~II.G in \cite{jackiw}.
This is dictated by the fusion rules of the bosonized Dirac--Weyl model
\cite{cgt}
\beqa
\ex^{\pm i\frac{1}{2}\phi'(z)} \ex^{\pm i\frac{1}{2}\phi'(w)} \sim 
(z-w)^{1/4}\ex^{\pm i\phi'(w)} 
\quad &\Longrightarrow& \quad
\gamma_F\left(\ex^{\pm i\frac{1}{2}\phi'}  \right)^2 =(-1) \nn 
\ex^{i\frac{1}{2}\phi'(z)} \ex^{-i\frac{1}{2}\phi'(w)} \sim 
(z-w)^{-1/4} 1 
\quad &\Longrightarrow& \quad
\gamma_F\left(\ex^{i\frac{1}{2}\phi'}  \right) 
\gamma_F\left(\ex^{- i\frac{1}{2}\phi'}  \right) = 1 \nn 
\ex^{i\phi(z)'} \ex^{-i\frac{1}{2}\phi'(w)} \sim 
(z-w)^{-1/2} \ex^{i\frac{1}{2}\phi'(w)} 
\quad &\Longrightarrow& \quad
\gamma_F\left(\ex^{-i\frac{1}{2}\phi'}  \right) = -
\gamma_F\left(\ex^{ i\frac{1}{2}\phi'}  \right). \nonumber 
\eeqa  
Therefore 
$\gamma_F\left(\ex^{\pm  i\frac{1}{2}\phi'}  \right) =(-1)^{\pm 1/2}$
and the fermionic parity can be consistently identified with
\beq\label{j_0}
\gamma_F=\ex^{i\pi j_0}, \quad j_0=\oint \frac{d\,z}{2\pi} j(z),
\quad j(z)= :\psi^*(z)\psi(z):= i \frac{\partial}{\partial z}\phi'(z).
\eeq
The Dirac--Weyl (neutral) current $j(z)$ in Eq.~(\ref{j_0}) generates
an additional $\u1$ symmetry present in the 331 model which is then
explicitly broken by adding an inter-layer tunneling term in
the Hamiltonian \cite{read-green}. When the tunneling becomes strong 
enough  there is a transition to the MR state, in which 
 the Dirac--Weyl characters
are reduced to the Majorana--Weyl ones \cite{cgt}, i.e., 
$K_{\pm 1}(\t,0;4)\to \ch_{1/16}(\t)$.
Since the chiral fermion 
parity in the R-sector is broken due to the non-abelian
fusion rules (\ref{fusion}), the MR
characters (\ref{chi_MR}) with $l=\pm 3$ are completely 
equivalent\footnote{the identification is done
according to the minimal value of the electric charge in the 331 
state, which is preserved during the projection to the Pfaffian state, 
i.e.,  $Q_{\min}=1/4$ for $\lambda= {-3},1$ and  $Q_{\min}= {-1/4}$ 
for $\lambda= {-1},3$ }
to those  with $l=\mp 1$, 
as the only quantum number distinguishing 
between them would be the fermion parity if implementable. 
Therefore  the terms with $l=\pm 3$ are excluded from the sum in 
Eq.~(\ref{Z_MR}), which gives $6$ topologically 
inequivalent quasiparticle excitations, i.e., 
the topological order becomes $6$ after the transition 
\cite{milo-read,cgt,read-green}. 
This fact is also well explained after Eq.~(4.9) in 
\cite{milo-read} where the untwisted sector topological
degeneracy $2q$ is replaced by $q$ in the twisted sector
\textit{``since the distinction between even and odd sectors
no longer applies"}.
The same could be concluded from the partition function Eq.~(4.20)  in 
\cite{milo-read}, which does not look diagonal in the characters
 due to the presence of fermionic zero modes that are treated 
separately. However,  after taking into account these 
zero modes,  the twisted characters become identical, i.e., 
$\chi_{(r+1/2)/q,ev,tw}^{331}=$$\chi_{(r+1/2)/q,od,tw}^{331}$.
While the twisted 331 characters in \cite{milo-read} can be labelled 
by the $\Z_2$ number even/odd, the corresponding twisted MR characters 
do not possess this number. 
Note that the $\Z_2$ parity number of \cite{milo-read} is not exactly 
the fermionic parity but is related to it and, more important,
the breaking of the former implies the breaking of the latter.
In other words, the spontaneous breaking 
of the chiral fermion parity is reflected by a reduction of the 
topological order and, vice versa,  
\textit{the reduction of the topological order in a transition between 
states with ``the same" symmetry is a signal of spontaneous
breaking of some (discrete) symmetry}.

The above mentioned spontaneous breaking could be understood 
algebraically  as follows.
The representation of $\gamma_F$
in terms of creation and annihilation operators is non-local and 
depends on the super-selection sector, i.e., on the boundary conditions
in the spatial direction. In the Neveu--Schwarz (NS)
sector \cite{CFT-book}, which is characterized by anti-periodic 
boundary conditions  for the Majorana field on the cylinder 
$\varphi(\theta+2\pi)=-\varphi(\theta)$  and periodic ones  
on the conformal plane  \cite{CFT-book}
$\varphi(\ex^{2\pi i}z)=\varphi(z)$, the fermionic parity is given by
\[
\gamma_F =(-1)^{\sum\limits_{n=1}^{\infty} \varphi_{-n+\frac{1}{2}} 
\varphi_{n-\frac{1}{2}}},
\quad \mathrm{where} \quad 
\varphi(z)=\sum_{n\in\Z}\varphi_{n-\frac{1}{2}}\ z^{-n}
\quad (\mathrm{NS-sector})
\]
is the mode expansion of the Majorana fermion. Then 
the property $\gamma_F |0\ra =|0\ra $ in the NS sector 
follows from the fact that the positive  modes of $\varphi$ 
annihilate the vacuum, i.e., 
$\varphi_{n-1/2}|0\ra =0$ for $n\ge 1$.
However, in the Ramond (R) sector \cite{CFT-book}, 
in which the Majorana fermion
is periodic on the cylinder $\varphi(\theta+2\pi)=\varphi(\theta)$ 
but anti-periodic on the conformal plane, i.e., 
\beq\label{fi_R}
\varphi(z)=\sum_{n\in\Z}\varphi_{n}\ z^{-n-\frac{1}{2}},
\quad \varphi(\ex^{2\pi i}z)=-\varphi(z)
\quad (\mathrm{R-sector}),
\eeq
the situation is more complicated due to the presence of a fermionic
zero mode whose square is not $0$. The anti-commutation relations 
$\{\varphi_n,\varphi_m \}=\delta_{n+m,0}$ imply that 
$(\varphi_0)^2=1/2$, which ultimately leads to a spontaneous 
breaking of the $\gamma_F$ symmetry in the R-sector.
Indeed, the two operators $\gamma_F$ and 
$\varphi_0$ form a $2$-dimensional Clifford  algebra
\beq\label{cliff}
\{\gamma_F,\varphi_0 \}=0, \quad (\varphi_0)^2=1/2, \quad
\gamma_F^2=1,
\eeq
whose lowest dimensional representation is given by the Pauli 
matrices $\s_i$. 
This means that the lowest-weight vector in the R-sector
should be double degenerated
and choosing a $\gamma_F$-diagonal basis of (orthogonal) ground states
$|+\ra$ and $|-\ra =\sqrt{2} \varphi_0 |+\ra$, we can write \cite{ginsparg}
\[
\gamma_F=\sigma_3 \ 
(-1)^{\sum\limits_{n=1}^{\infty} \varphi_{-n}  \varphi_{n}},
\quad \varphi_0= \frac{1}{\sqrt{2}} \sigma_1 \
(-1)^{\sum\limits_{n=1}^{\infty} \varphi_{-n}  \varphi_{n}}\quad 
(\mathrm{R-sector}).
\]
Put another way, \textit{the fermion parity $\gamma_F$ in the MR state 
does not
have  $1$-dimensional representations (unlike the 331 state) 
due to the presence of a fermionic
zero mode whose square is not $0$}.

On the other hand, the double degeneration of the (neutral) vacuum 
subspace in the R-sector of the MR state  
is  incompatible with modular invariance. The point is that 
the modular invariance 
of systems like the Ising model requires a projection on states with a 
given value of $\gamma_F$, which
is called the GSO projection in string theories \cite{ginsparg}.
Retention of (orthogonal) states with both parities in the R-sector of 
the Ising model necessarily breaks  modular  invariance \cite{ginsparg}. 
This could also be checked directly since if we consider two sectors 
with the same (neutral part) minimal CFT dimension $1/16$ 
(and opposite parities) we have to 
put a factor of $2$ in front of the
character $|\chi_{\pm 1}(\t,\z)|^2$ in Eq.~(\ref{Z_MR}) and this 
new modular matrix  $N_{ij}$ \cite{CFT-book} does not commute with the 
MR $S$-matrix \cite{cgt}.
Moreover,  this doubling is physically irrelevant since the factor 2,
which is present in the 331 partition function \cite{milo-read}, 
comes from the number of 
independent fermionic zero modes in the 331 model \cite{milo-read}.
While there are 2 independent  zero modes in the 331 model, 
$\psi_0^*$ and  $\psi_0$,  representing pseudo-spin up and down 
respectively,   only one, $\varphi_0=(\psi^*_0+\psi_0)/\sqrt{2}$  
survives the projection \cite{cgt} $\psi^*-\psi\to 0$, which implements 
the high-tunneling limit from the 331 to the Pfaffian state, i.e., 
there is only one such a mode in the Pfaffian state.
Note that the   modular invariance is of fundamental importance for the 
FQH effect effective field theories, so we consider only one ground
state in the R-sector \cite{milo-read,cgt}, 
like in the partition function (\ref{Z_MR}).  

Thus, we conclude that 
\textit{the necessity to choose  exactly one ground state in the 
R-sector of the Ising model (GSO projection), together with the fact 
that $\gamma_F$ does not have  one-dimensional representations in this 
sector,  results in the spontaneous breaking of chiral fermion parity 
in the twisted sector of the Pfaffian model}.

Therefore the fusion rules (\ref{fusion}) of the chiral spin 
fields in the Ising model cannot preserve any fermion
parity number as noted in Sect.~\ref{sec:mr}. 
We stress that the chiral fermion parity is crucial 
for the FQH states due to the chiral nature of the edge states. 
In particular, according to the Leggett conjecture \cite{leggett}, 
the FQH ground states are expected to be paramagnetic for $\gamma_F=+1$
and diamagnetic for $\gamma_F=-1$.
We note that this problem does not exist in the EPf state since 
the  R-sector is trivial there.
\section{PH transformation and PH symmetry}
\label{sec:PH}
The PH symmetry of the wave functions describing FQH  
states  is a reflection of the charge conjugation  (or $C$-) symmetry 
of the  topological Chern--Simons field theories, which are known 
\cite{fro2000} to be the large-scale/low-energy effective field theories 
for the bulk of the incompressible $2+1$ dimensional  electron systems. 
Therefore the $C$-symmetry is supposed to be  
an exact symmetry of all QH states 
in the thermodynamic limit (cf.  \cite{jain-goldman} and \cite{girvin}) 
and should preserve the universality class (but not the non-universal 
quantities such as the energy gap).
In particular, the MR and the EPf state are expected to be PH-symmetric,
as is easily seen from the partition functions (\ref{Z_MR}) and  
(\ref{Z_EPf}).
However, the numerical calculations \cite{5-2hr}, 
which show that the MR state is most likely  the true ground state 
for $\nu=5/2$ at $T=0$, start from a model Hamiltonian which is not 
PH symmetric.
This is confusing since  the PH symmetry must be a characteristics of the 
universality class and two Hamiltonians with different PH symmetry 
cannot be equivalent. In order to explain this we propose the following 
interpretation.
Let us consider two ground states (GS) $|+\ra$ and 
$|-\ra$ of a PH symmetric Hamiltonian,  which are respectively even and 
odd under parity.  Then we can construct the linear combination
$|\MR\ra = (|+ \ra + |-\ra )/\sqrt{2} $, which has no definite PH parity 
but the same energy as $|+ \ra$ and $|- \ra$ and probably 
might be determined as the ground state of some other (trial)
Hamiltonian.
The fact that the (true) Hamiltonian for $\nu=5/2$ 
commutes with the PH transformation means
that its eigenstates form representations of the $\Z_2$ group, i.e.,
they are either even or odd and both  have the same energy. 
Therefore the unique  exact  ground state $|\NUM \ra$  for
$\nu=5/2$   is  PH symmetric \cite{5-2hr}. 
Next, motivated by the results in \cite{5-2hr}, we assume 
that the exact ground state is 
$| \NUM \ra = | + \ra$  and compute the 
overlap with the  GSs  $|+ \ra$, $|-\ra$ to be  $\la \NUM |+ \ra = 1$ and 
$\la \NUM | - \ra=0$, 
which is natural and consistent with the preservation of the 
PH symmetry. Then, the projection $\vert\la\NUM | \MR \ra\vert$ would be
$1/\sqrt{2}\simeq 0.71$, while that for the symmetrized MR would be
$\vert\la\NUM | P_+ |\MR \ra\vert =1$. 
The results of \cite{5-2hr} are that the projection 
 $\vert \la\NUM | \MR \ra \vert$ of the MR state on the exact GS
does not exceed $73\%$, 
while that of the ``PH symmetrized MR" is $97 \%$, which are in good
agreement with $0.71$ and $1$ respectively. We expect that the difference
between the estimated projections  and those computed with $N=10$ 
electrons \cite{5-2hr} would decrease when $N$ increases.
Note that, in view of the above interpretation,
 the PH-symmetrized MR state is expected to be 
equivalent to $|+\ra$, i.e.,  
to $| \NUM \ra$ since the PH symmetrization is implemented by the 
projector $P_+=|+\ra \la +|$. 
\section{Spin--charge separation and specific heat in the EPf state}
\label{sec:heat}
In Sect.~\ref{sec:epf} we have shown that the partition function of
the EPf state is invariant under $T^2$, $S$, $U$ and $V$ 
transformations. The covariance of the CFT characters under $S$
guarantees that the quasiparticle spectrum is closed and complete,
as explained after Eq.~(\ref{Z_EPf}).
Another necessary completeness check for this spectrum comes from the 
fact that the central charge of the CFT determines the 
Casimir free energy on the cylinder \cite{CFT-book} and therefore
the specific heat contributions \cite{elburg-schout} 
$\gamma_{\mathrm{q.p.}}$ of all independent quasiparticle excitations 
should sum up to $\gamma_{\mathrm{CFT}}=\frac{\pi}{6} c$, where $c$ is the 
CFT central charge.  
The corresponding computation for the MR state can be found 
in \cite{schout2}.

The complete spin--charge separation in the EPf state, 
discussed after Eq. (\ref{EPf-q.h.}), 
allows us to choose the following complete
set of independent quasiparticle excitations for the EPf state
\[
\left\{
\varphi(z), \quad :\ex^{-i\sqrt{2}\phi(z)}:, \quad
:\ex^{i\frac{1}{\sqrt{2}}\phi(z)}:
\right\} .
\]
Using this basis we can  compute the specific heat
contribution in the framework of the exclusion statistics
\cite{elburg-schout}
\[
\frac{C_g}{L}=-k_B \beta^2 \frac{\partial}{\partial\beta}
\rho_0 \int_{0}^\infty  d  \epsilon \, \epsilon \,
n_g(\epsilon-\mu) = \rho_0 \gamma_g   k_B^2 T ,
\]
where $L$ is the length of the edge and $\r_0=(\hbar v_F)^{-1}$ is the
density of states per unit length \cite{elburg-schout}
($v_F$ is the  Fermi velocity on the edge).
Integrating the anyon distributions
$n_g$ for $g=1$, $2$ and $1/2$ (cf. \cite{elburg-schout}), we find
$\gamma_\varphi=\gamma_1=\frac{\pi}{6}\frac{1}{2}$,
$\gamma_2=\frac{\pi}{6}\frac{2}{5}$,
and  $\gamma_{1/2}=\frac{\pi}{6}\frac{3}{5}$,
 which give a total specific heat coefficient
$\gamma= \gamma_1+ \gamma_2+ \gamma_{1/2} =\frac{\pi}{6}\frac{3}{2}$,
exactly reproducing the central charge of the edge CFT,
as it should be.
\section{Chiral persistent currents}
\label{sec:pers}
According to a (well-known but unpublished) Bloch theorem, 
the free energy of a conducting ring (or other non-simply connected 
conductor) is a periodic function of the magnetic flux through the 
ring with period one flux quantum $\phi_0=h/e=1$. 
The flux dependence of the free energy 
$F(T,\phi)=-k_B T \ln(Z(T,\phi))$ within one period 
gives rise to  an equilibrium current 
$I=-(e/h) (\partial / \partial \phi)   F(T,\phi) $, called a 
\textit{persistent current},
flowing along the ring, which has a universal amplitude 
and temperature dependence.
Such currents have been  observed in mesoscopic rings \cite{pers-exp},
where the length of the ring is smaller than the coherence length
so that the electronic transport is coherent. 

The persistent currents, flowing along the edge of a mesoscopic 
disk FQH sample,  can give important information about the 
low-temperature behaviour of the FQH states 
\cite{geller-loss-kircz,ino}.
Here we stress again the advantage of the CFT approach 
in the computation of the oscillating persistent currents
as compared to
the chiral Luttinger liquid description of the edge states
which is rather unsatisfactory for general filling factors \cite{grayson}.
In order to compute the persistent current for a single FQH edge,
that is more relevant for the  experiment,  
we need to implement the process of threading the sample by 
a (fractional) flux. To this end we consider 
the \textit{chiral partition function},
i.e., the linear partition function 
\beq\label{Z_chi}
Z^+(\t,\z)=\sum\limits_{l} \chi_{l}(\t,\z)
\eeq
constructed for one of the edges (right-moving here)
\beqa\label{Z_chi_EPf}
Z_{\EPf}^+(\t,\z)&=&\ex^{-\frac{\pi}{2}\frac{\left(\Im\z\right)^2}{\Im\t}}
\Bigl( K_0(\t,\z;2)+K_1(\t,\z;2)\Bigr) 
\left(\ch_0(\t)+\ch_{1/2}(\t) \right) 
\eeqa
\beqa\label{Z_chi_MR}
Z_{\MR}^+(\t,\z)&=&\ex^{-\frac{\pi}{2}\frac{\left(\Im\z\right)^2}{\Im\t}}
\Bigl( K_0(\t,\z;2)+K_1(\t,\z;2)\Bigr) 
\left(\ch_0(\t)+\ch_{1/2}(\t) \right) + \nn
&+&\ex^{-\frac{\pi}{2}\frac{\left(\Im\z\right)^2}{\Im\t}}
\left( K_{1/2}(\t,\z;2)+K_{-1/2}(\t,\z;2)\right) 
\ch_{1/16}(\t), 
\eeqa
where the characters  $\chi_l(\t,\z)$ for the MR and EPf states are 
given by  Eq.~(\ref{chi_MR}) and Eq.~(\ref{chi_EPf}) respectively, and 
the factor in front of the sum is the crucial CZ factor
(\ref{CZ}) (we have also used the identity (\ref{K_2l})).
To realize the flux threading procedure, we note that since the 
$V$ transformation, Eq.~(\ref{V}), 
was interpreted as increasing the flux by one unit,
the transformation $\z\to\z+\phi \t$ means increasing the flux 
by $\phi$ in units $\phi_0=1$. Therefore,
the equilibrium \textit{chiral persistent current} in the FQH system 
can be computed directly from the CFT partition function by
\beq\label{pers}
I=\left(\frac{e}{h}\right) k_B T \frac{\partial}{\partial \phi}
\ln Z^+(\t,\z+\phi\t) , \quad \mathrm{where}\quad 
\t=i\pi \frac{T_0}{T}, \quad T_0=\frac{\hbar v_F}{\pi k_B L},
\eeq
$v_F$ is the Fermi velocity and $L=2\pi R$ the circumference of the 
edge.  
We note that the persistent current for the EPf state coincides with 
that for the chiral Luttinger liquid (or bosonic Laughlin quantum Hall 
state with $\nu=1/2$) as shown in Appendix~\ref{app:laugh}.

\begin{figure}[htb]
\centering
\caption{Persistent currents in the MR and EPf states
computed numerically for $T/T_0=0.1$. 
The flux is measured in units $h/e$ and the current's unit is 
$e v_F/4L$. 
The period is 1 flux quantum for both states and the amplitudes are
$I_\MR^{\max}=0.828$ and $I_\EPf^{\max}=0.887$.
\label{fig:pers}}
\epsfig{file=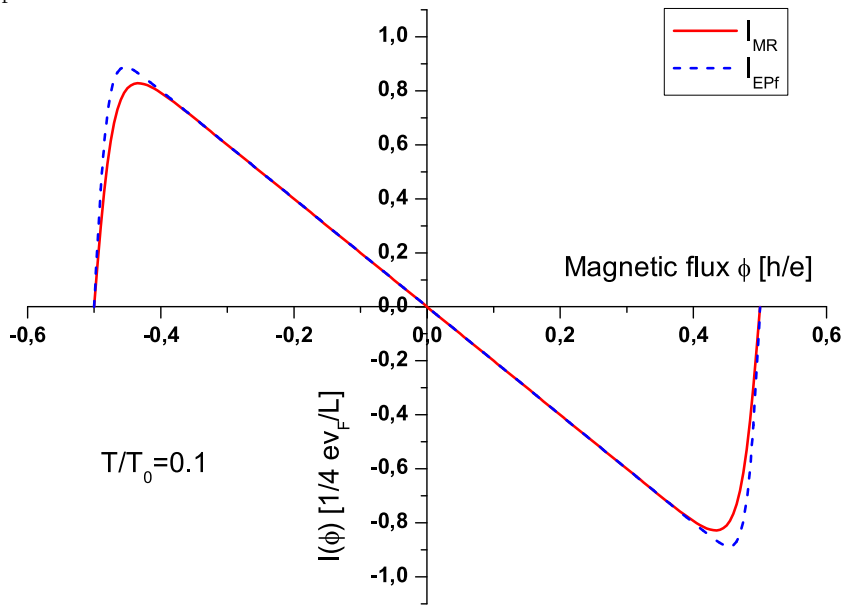,height=8cm}
\end{figure}

The partition functions (\ref{Z_chi_EPf}) and
(\ref{Z_chi_MR})   constructed for the  MR and EPf state
are invariant under the $V$ transformation  Eq.~(\ref{V}), which  
implies that the corresponding  chiral persistent currents should be 
periodic in $\phi$ with period at most  $1$.
The plots of the two chiral persistent currents 
computed numerically from Eq.~(\ref{pers}) 
for  $-1/2 \leq \phi \leq 1/2$ in the EPf and the MR 
states at temperature $T/T_0=0.1$, given on Fig. \ref{fig:pers},
\noindent
indicate that 
\textit{both currents are periodic in $\phi$ with period exactly $1$}.
We  do not see any anomalous oscillations, such as 
half-flux periodicity of the persistent 
currents\footnote{under certain conditions the period of the persistent 
currents for the paired states could be shown to be $1/2$ \cite{ino3}} 
\cite{ino,ino2}, which is 
characteristic for the BCS paired condensates,
or more generally for some broken symmetries, 
neither in the MR nor in the EPf states for all temperatures 
$0.05\leq T/T_0 \leq 9$ that we could access numerically.
This is exactly the content of the Bloch theorem, which is also known as 
the Byers--Yang theorem \cite{byers-yang} in the context of 
superconductors.

\begin{figure}[htb]
\centering
\caption{Temperature decay of the persistent current's amplitudes  
in the MR and EPf states.
The amplitude are computed numerically in units $e v_F/4L$,
the temperature is measured in units of $T_0$. \label{fig:decay}}
\epsfig{file=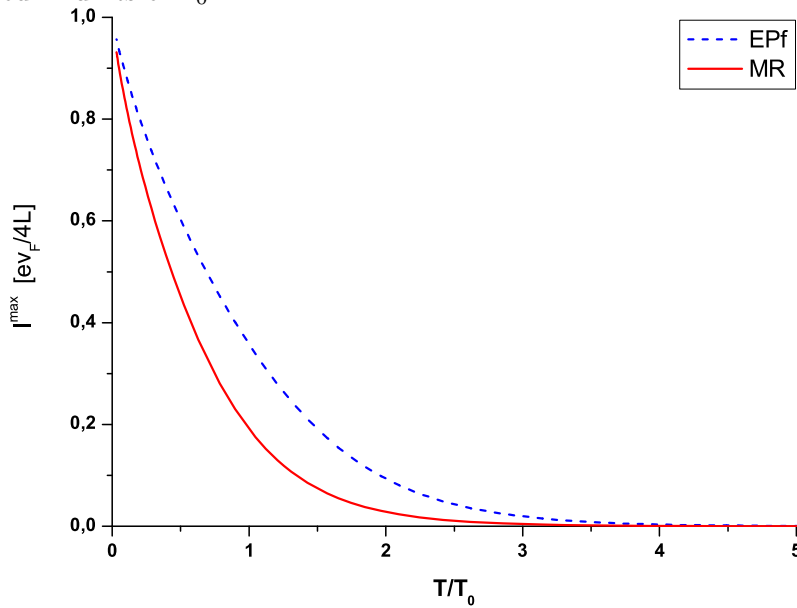,height=8cm}
\end{figure}

\begin{figure}[htb]
\begin{center}
\caption{Logarithmic plot of the  temperature dependence of the amplitudes
of the persistent currents (in units of $ev_F/4L$) in the MR and EPf 
states computed numerically
for temperatures $0.05\leq T/T_0 \leq 9$ \label{fig:logplot}}
\epsfig{file=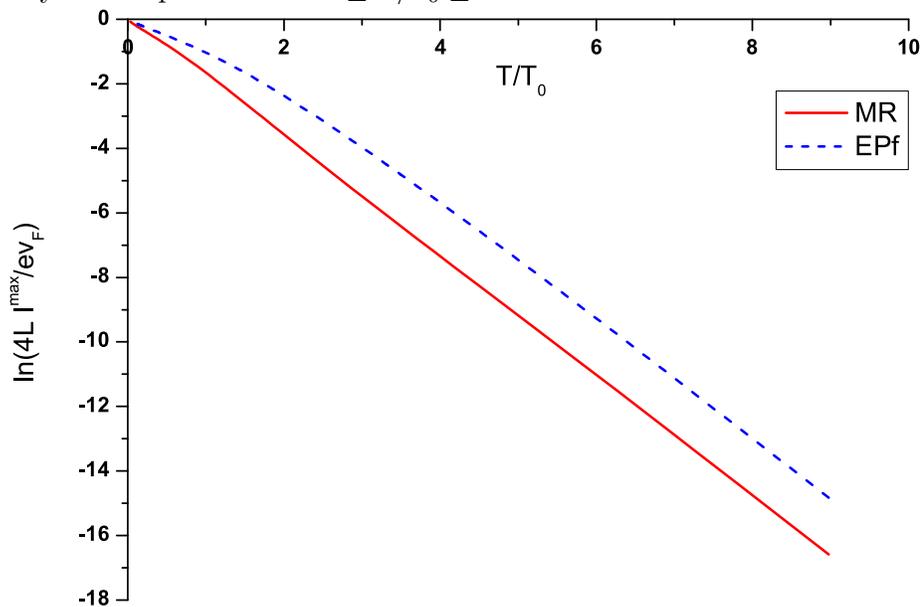,height=8cm}
\end{center}
\end{figure}

The amplitudes of the persistent currents in the EPf and MR 
states decay exponentially with temperature, as shown 
in Fig.~\ref{fig:decay}.
While at $T=0$ both currents have the same amplitude 
$I^{\max}_{\MR}=I^{\max}_{\EPf}$,
for finite temperature \textit{ the amplitude of 
the current  $I_\EPf$ is always  bigger than that of  $I_\MR$}.
This fact could be used to detect (e.g., in a SQUID experiment) any 
transition between the two states.
For convenience we also show 
on Fig.~\ref{fig:logplot} the logarithmic  plot of  the 
temperature decay Fig.~\ref{fig:decay} of the persistent currents' 
 amplitudes. 
The temperature dependence on Fig.~\ref{fig:logplot} 
is not linear, showing two distinct regions, 
which we shall conventionally call low-temperature and high-temperature 
regions  and shall investigate separately. 
This suggests the existence of  two different mechanisms reducing
the amplitude of the persistent currents at finite temperature.
We recall that the chiral persistent currents computed here are
universal in the sense  that due to the absence of 
backscattering from impurities, 
there is no reduction from weak disorder \cite{geller-loss} 
(for FQH states without simultaneously counter-propagating modes 
 \cite{michael}).
\subsection{Low temperature limit}
In the low temperature limit $T/T_0\to 0$ the modular parameter
vanishes, $q=\ex^{-2\pi^2 \frac{T_0}{T}}\to 0$, and  the low temperature 
asymptotics of the persistent current is determined by the leading term,
coming from the vacuum sector, which is proportional to  $q^0$ ,  
multiplied by the CZ factor (\ref{CZ}). Therefore
the partition functions  in the EPf and MR states 
are the asymptotically same in this limit
\[
Z^+_\EPf(T,\phi)=Z^+_\MR(T,\phi) \mathop{\simeq}\limits_{T/T_0\to 0}
\exp\left(-\frac{\pi^2}{2} \frac{T_0}{T} \phi^2\right).
\]
and the zero-temperature amplitudes of the persistent currents
in the EPf and MR states are given by (without the contribution from
the lowest Landau level)
\beq\label{0T}
I^{\max}_\EPf=I^{\max}_\MR = \frac{1}{4}\frac{e v_F}{L}
\quad \mathrm{for} \quad T=0.
\eeq
Note that in general there is a large, non-mesoscopic 
contribution to the zero temperature amplitude of the current
\cite{michael}, 
which is  proportional to the cyclotron frequency $\omega_c$, 
hence  slowly varying with $B$
(see Eq.~(37) in \cite{geller-loss-kircz}). This component of the 
persistent current 
cannot be derived within the CFT approach because the latter is
only a theory of the low-lying excitations above the ground state,
while the  $\omega_c$ contribution reflects the 
properties of the ground state, and involves states deep below the 
Fermi energy \cite{michael}.
Nevertheless, the oscillating part of the persistent currents derived 
here seems to be measurable in SQUID experiments \cite{pers-exp}.

\begin{figure}[htb]
\begin{center}
\caption{Low-temperature dependence  of the persistent currents' 
amplitudes  in the MR and EPf states computed numerically for 
$0.05 \leq T/T_0 \leq 0.3$ and analytically, using Eqs. (\ref{low-T-EPf})
and (\ref{low-T-MR}), for $0 \leq T/T_0 \leq 0.3$. \label{fig:low-T}}
\epsfig{file=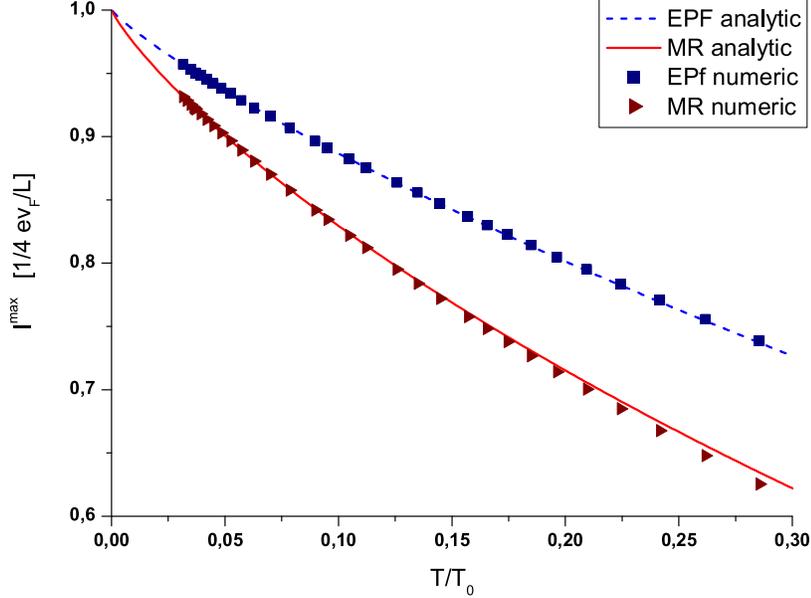,height=8cm}
\end{center}
\end{figure}

Taking into account also the next-to-leading order contribution to 
the mesoscopic persistent current $I(T,\phi)$ and finding its maximum 
for $T$-fixed gives the following low-temperature asymptotics for the 
current's amplitude in the EPf state
(see Appendix~\ref{app:low-T} for details)
\beqa\label{low-T-EPf}
I^{\max}_\EPf(T) & \quad  \mathop{\simeq}\limits_{T/T_0<<1} \quad &
\frac{e v_F}{L} \frac{1}{2\pi^2}\frac{T}{T_0} 
\mathrm{arccosh}\left( \frac{1}{2\pi^2}\frac{T}{T_0} 
\exp\left(\frac{\pi^2}{2}\frac{T_0}{T}\right)  \right) - \nn  
& \quad  - \quad &
\frac{e v_F}{L} \left[\left(\frac{1}{2\pi^2}\right)^2 
\left(\frac{T}{T_0}\right)^2 - 
\exp\left(-\pi^2 \frac{T_0}{T} \right) \right]^{1/2}
\eeqa
and 
\beqa\label{low-T-MR}
I^{\max}_\MR(T) & \quad  \mathop{\simeq}\limits_{T/T_0<<1} \quad  &
\frac{e v_F}{L} \frac{1}{\pi^2}\frac{T}{T_0} 
\mathrm{arccosh}\left( \frac{2}{\pi^2}\frac{T}{T_0} 
\exp\left(\frac{\pi^2}{4}\frac{T_0}{T}\right)  \right) - \nn  
& \quad  - \quad &
\frac{1}{2}\frac{e v_F}{L} \left[\left(\frac{2}{\pi^2}\right)^2 
\left(\frac{T}{T_0}\right)^2 - 
\exp\left(-\frac{\pi^2}{2} \frac{T_0}{T} \right) \right]^{1/2}
\eeqa
for that in the MR state.
The low-temperature region, corresponding to $0\leq T/T_0 \leq 0.3$, 
is shown separately on Fig.~\ref{fig:low-T}. 
We note that, according to Fig.~\ref{fig:low-T},  
the asymptotic formulae (\ref{low-T-EPf}) and (\ref{low-T-MR}) 
give excellent approximations for $T/T_0\leq 0.1$.

The mechanism responsible for this decay \cite{pers-exp}
is the mixing of  contributions of energy levels in an energy interval 
$k_B T$, which reduces the current since adjacent levels give opposite
contributions. The characteristic scale for this mixing is given
by the energy gap (not simply  by the level spacing $\hbar \ 2\pi v_F/L$). 
\subsection{High temperature limit}
The high-temperature limit $T/T_0\to \infty$ is a non-trivial one
since $q=\ex^{-2\pi^2 \frac{T_0}{T}}\to 1$ is at the border of the 
convergence interval for the partition functions. However, since the
latter is constructed as a sum of RCFT characters, one could use their 
$S$-covariance to relate the high-temperature and low-temperature limits
(in proper modular parameters).
Note that, unlike the annulus partition functions, the disk partition 
functions (\ref{Z_chi_EPf}) and (\ref{Z_chi_MR}) are not $S$-invariant 
and as we shall see this leads to a completely different temperature 
behaviour after $S$-transformation.  We find that the amplitudes of the 
persistent currents in the EPf and MR states decay exponentially
with the temperature
(see Appendix~\ref{app:high-T} for details)
for $T/T_0 >>1$, i.e.,
\beq\label{high-T-EPf}
I^{\max}_\EPf (T) \quad \mathop{\simeq}\limits_{T/T_0 >>1} \quad
\frac{2}{\pi} \frac{e v_F}{L} \left( \frac{T}{T_0}\right)
\exp\left(-2\frac{T}{T_0}\right)
\eeq
\beq\label{high-T-MR}
I^{\max}_\MR (T) \quad \mathop{\simeq}\limits_{T/T_0 >>1} \quad
\frac{2}{\pi} \frac{e v_F}{L} 
\left(\frac{\sqrt{2}-1}{\sqrt{2}+1}\right)
\left( \frac{T}{T_0}\right)
\exp\left(-2\frac{T}{T_0}\right)
\eeq
The logarithmic plots of the amplitudes of the persistent currents in 
the EPf and MR states, computed numerically, 
in the high-temperature region, corresponding  to $1\leq T/T_0\leq 9$ 
which are given in Fig.~\ref{fig:logplot}
become almost linear, with the same slopes but different $y$-intercepts.
After removing the subleading $\ln(T/T_0)$ contribution,
which is explicit in Eqs.~(\ref{high-T-EPf}) and (\ref{high-T-MR}),
the Least-Squares fit gives  $-2.00005$ for the slope in the EPf state 
and $-2.004$ for 
that in the MR state, while the $y$-intercepts   
give amplitudes (in units $e v_F/4L$) $2.547$ for the EPf 
and $0.452$ for the MR states 
respectively.
Again, we see that the high-temperature asymptotic formulae 
(\ref{high-T-EPf}) and (\ref{high-T-MR}) are in excellent agreement
with the numerical calculations shown in Fig.~\ref{fig:logplot}.
The ratio $I^{\max}_\EPf /I^{\max}_\MR $ increases to $5.813$ 
for $T/T_0 \to 9$, which is very close to the analytic value 
$(\sqrt{2}+1)/(\sqrt{2}-1)\simeq 5.828$, that gives the universal 
ratio for $T/T_0\to\infty$.

This high-temperature universal non-Fermi 
liquid  behaviour, characterized by the 
temperature $T_0$, which is closely related to the level spacing, 
expresses another mechanism for  
``thermal smearing" \cite{michael}  of the persistent currents. 
\section{Phase transition between the EPf and MR states}
\label{sec:PT}
Although  continuous symmetries in (1+1) D cannot be spontaneously
broken \cite{coleman}, there could be a spontaneous
breaking of some discrete symmetry
at finite temperature $T_c>0$.
As we have shown in Sect.~\ref{sec:parity} the $\Z_2$ fermion parity is
spontaneously broken in the R-sector of the Ising model.
Therefore we believe that there is a classical II-nd order 
 phase transition, at $\nu=5/2$, 
which  is characterized by the spontaneous breaking of this $\Z_2$ 
symmetry. The high temperature ``symmetric" phase
($T>T_c$) corresponds to the EPf state, in which the chiral fermion 
parity is well-defined, while in the low temperature ``ordered" phase
($T<T_c$), corresponding to the MR state,  this symmetry is spontaneously
broken.
This is different from the spontaneous magnetization in the 
2D Ising model where the $\Z_2$ symmetry changes the sign of the
spin field $\s\to -\s$, while leaving the fermion invariant.
In the present case the generator  $\gamma_F$ of the symmetry
anticommutes with the fermion $\varphi(z)$
and the ``order parameter", which is given by the vacuum expectation 
value $\la \varphi(\theta) \ra_{\mathrm{PBC}}$ in the sector with periodic 
boundary conditions on the cylinder, 
is fermionic and hence not directly observable.
Nevertheless it has crucial
implications for the corresponding phases, such as change of the ground 
state energy and spectrum of excitations  on the cylinder and  
diamagnetic/paramagnetic ground state's structure 
depending on the fermion parity.   

The nature of this phase transition is not clear despite the
extensive numerical work \cite{morf,5-2hr}.
Motivated by the analysis of the chiral fermion parity,
we propose the following scenario: as temperature increases
the system undergoes a II-nd order phase transition, from the  
MR state, in which the chiral fermion parity is spontaneously broken,
 to an intermediate compressible Composite Fermions (CF) Fermi liquid 
 state found before \cite{morf,5-2hr,fermi5-2}, which possesses 
this $\Z_2$ symmetry, 
and then  a I-st order phase transition from
the CF state to the EPf one.
Since the resistance $R_0$ in Fig.~\ref{fig:activ} has a jump,
one would expect that the transition MR $\to$ EPf is simply of 
first order. 
However, according to our analysis, this transition is
accompanied by spontaneous breaking of the $\Z_2$ chiral fermion parity
 symmetry discussed in Sect.~\ref{sec:parity}. On the other hand, 
a second order transition between the EPf and the MR states,
with spontaneous breaking of the fermion parity,  
cannot occur directly since 
incompressible Hall states with different topological orders,
such as the MR and the EPf states,  support
residual neutral gapless modes at the phase boundary,
leading to a first order transition \cite{mcd-hald,2-3nay-wilc}.
However, there is a possibility for a two-step process involving an
intermediate compressible state \cite{mcd-hald,2-3nay-wilc}.
In that case, as temperature decreases, the EPf state undergoes 
a first order phase transition to the CF Fermi liquid state, 
in which the $\Z_2$ symmetry is preserved, and then a second order 
phase transition to the  MR state in which the $\Z_2$ symmetry is broken.

Note that, unlike the MR state, the compressible
CF state\cite{5-2hr}  is $\Z_2$ symmetric and, at the same time
(despite its compressibility) has the
topological structure  similar to that of the MR state.
Indeed, as pointed out in \cite{5-2hr}, the $4$ values of the total 
momentum $\bf{K}$ 
($3$ distinct values) correspond to the $3$ distinct values 
of $\bf{K}$ in the MR state on the torus. 
In fact, the CF state has the  topological structure of the 331 state, 
which possesses this $\Z_2$ symmetry, see Sect.~\ref{sec:parity}. 
Therefore we believe that the transition MR $\to$ CF is in the same 
universality class as the transition 331 $\to$ MR and is characterized 
by the spontaneous breaking of the chiral fermion parity.
We stress that due to the  similar  topological structure  of the MR 
 and CF states, the phase boundary between these phases do not support 
gapless neutral modes \cite{mcd-hald}, which opens the possibility 
of a second order transition, in which the fermion parity is 
spontaneously broken.
On the other hand, because of the topological  mismatch, the transition
 from the CF to the  EPf state  could only be of first order.
This scenario is in agreement with the numerical calculations
\cite{morf,5-2hr} as well as with the activation experiment
at $\nu=5/2$, see Fig.~3 in \cite{pan}, where a 
 ``kink" was  observed around $T=15$ mK.

\begin{figure}[htb]
\centering
\epsfig{file=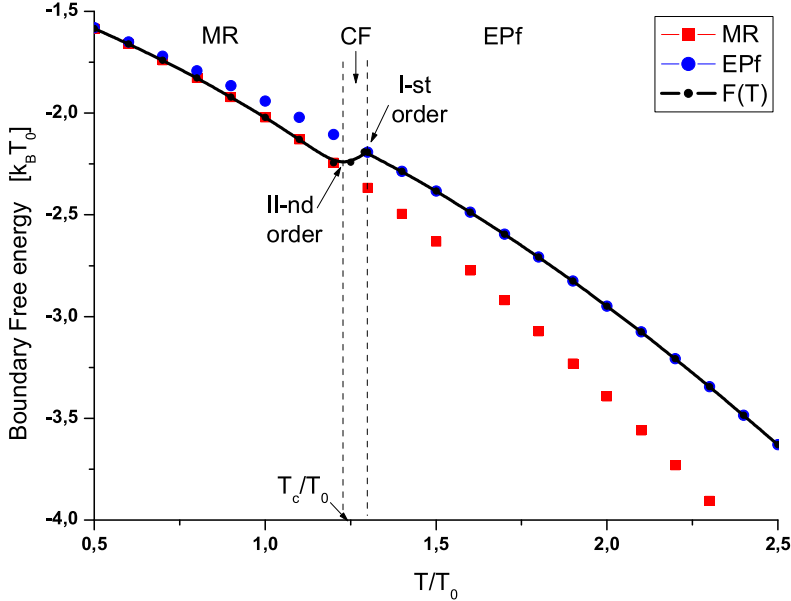,height=8cm}
\caption{Boundary free energy for the MR and EPf states at 
low temperature computed numerically from CFT
(without additional flux, i.e., $\phi=0$). The free energy 
for the intermediate compressible  phase, between the two vertical 
dashed lines,  is qualitative. The black line shows the expected
behaviour of the free energy $F(T)$ for $\nu=5/2$. 
\label{fig:PT}}
\end{figure}

\begin{figure}[htb]
\centering
\epsfig{file=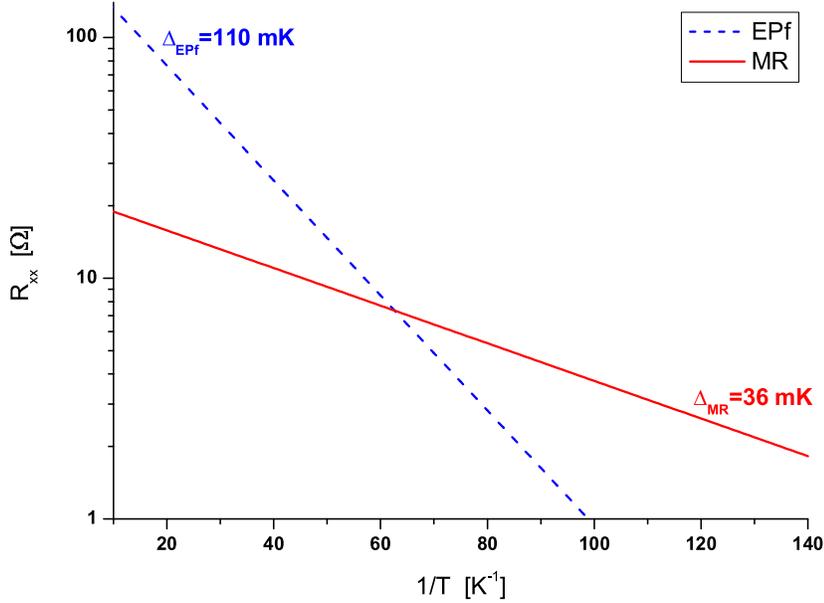,height=8cm}
\caption{Logarithmic plot of the longitudinal resistance 
in the MR and EPf states as a function of the inverse temperature.
The slopes (i.e., the gaps) for both states correspond to the sample 
of \cite{pan} and the change of the slope is an indication of a phase 
transition \label{fig:activ}}
\end{figure}

In Fig.~\ref{fig:PT} we plot the low-temperature behaviour of the 
free energy on the edge as computed numerically from the CFT. 
We stress that the CFT description of the FQH system is valid as long
as the system is incompressible, i.e., for  temperatures well below the
activation energy. 
Note that the CFT dimension, which is equal to the 
\textit{average} spin \cite{gaps},  is proportional to the (ideal) 
\textit{average} quasiparticle--quasihole energy \cite{gaps}, i.e., 
the activation energy is a half of the gap computed in 
Eq.~(\ref{gap_EPf}) and Eq.~(\ref{gap_MR}).  
Therefore in systems such as the $\nu=5/2$, 
with more than 
one phases with different gaps, there might be an interesting interplay
between the various phases at different temperatures.
As can be seen from Fig.~\ref{fig:PT}, the free energy  of the MR state 
is always lower than that of 
the EPf and  we believe that the FQH system is in the MR phase for 
temperature $ T< \D_\MR /2 k_B $, i.e., until the  
free energy for the MR state computed from the CFT is a good 
approximation. 
This is confirmed by numerical
calculations at zero temperature \cite{5-2hr}.
The free energy for the MR and EPf states  at zero temperature
are the same, i.e.,  $F_{\MR}(T=0)=F_{\EPf}(T=0)$, as it should be since 
both phases share the same absolute ground state.
For temperature higher than the MR activation energy $T>\D_\MR/2 k_B $, 
the boundary free energy goes very fast to its  
value at zero temperature (the ground state energy)
since 
the lowest-energy charged edge excitations are quickly transferred 
into the bulk
where they have less momentum and hence less energy as compared to the 
edge. This may lead to a  II-nd order  phase transition, 
 at  critical temperature $ T_c\simeq \D_\MR / 2 k_B $ characterized  
by  the energy  gap in the MR state,
 from the MR state to a compressible state, which 
seems to be topologically equivalent to the CF Fermi liquid state.
Then, for temperature very close to $T_c$, the free energy of the 
intermediate phase CF crosses the free energy of the EPf state 
leading to a I-st order phase transition CF $\to$ EPf as shown in
Fig.~\ref{fig:PT}. This is consistent 
with the experimental observation of the change in the slope of 
$\ln(R_{xx})$, which looks like a single transition.
The characteristic temperature for this two-step transition
in the absence of disorder 
could be estimated using the gap ansatz
(\ref{gap})  with $\Delta_{\mathrm{q.h.}}^{\MR}=1/8$ for 
$\Gamma=0$ and $\alpha$ being the ratio of the level 
spacing and Coulomb energy, to be
\beq\label{T_c}
T_c=\frac{\D_{\MR}}{2 k_B}=
\pi^2 T_0 \Delta_{\mathrm{q.h.}}^{\MR}  = \frac{1}{2}
\hbar \frac{2\pi v_F}{L} \Delta_{\mathrm{q.h.}}^{\MR} \quad
\Longrightarrow \quad \frac{T_c}{T_0}=\frac{\pi^2}{8}.
\eeq
The different gaps, Eqs.~(\ref{gap_EPf}) and (\ref{gap_MR})
for the sample of \cite{pan}, 
 in both phases lead to a change of the slope in the logarithmic plot
 of the diagonal resistance
$R_{xx}=R_0 \exp(-\D/2k_B T)$,
as a function of $1/T$, in the thermal activation experiment, 
which is illustrated on  Fig.~\ref{fig:activ}.
The trial values of $R_0^\EPf=230 \Omega$, $\D_\MR=36$mK and 
$R_0^\EPf/R_0^\MR\simeq 10.2$, which were used to plot 
Fig.~\ref{fig:activ},
 have been chosen for maximal overlap
with Fig.~3 in \cite{pan}. Our fit shows that the 
temperature at which the two lines 
$\ln R_{xx}^\EPf=\ln R_0^\EPf-\D_{\EPf}/2k_B T$ and 
$\ln R_{xx}^\MR=\ln R_0^\MR-\D_{\MR}/2k_B T$ 
intersect is not exactly $T=\D_\MR/2k_B \simeq 18$mK but rather 
$T=\D_\MR/2k_B -\delta T$, where the shift $\delta T$ is determined
from
\[
\frac{R_0^{\EPf}}{R_0^{\MR}}=
\exp\left(\frac{\D_{\EPf}- \D_{\MR}}{\D_{\MR}-2\delta T}\right)
\simeq 10.2 \quad \Longrightarrow \quad \delta T\simeq 2 \mathrm{mK}.
\]
This might be an indication that the transition MR $\to$ EPF is not 
simply of I-st order, but involves an intermediate state.
When the residual disorder is taken into account, 
the critical temperature (\ref{T_c}) is supposed to 
decrease in the same way 
like the energy gap \cite{gaps}. For example, our scenario implies 
for the sample of \cite{pan}, where $\D_\MR\simeq 36$ mK,  
that $T_c\simeq 16$ mK and for $T<T_c$ (i.e., for $1/T>62$) 
the system is in the MR phase, for $16 < T < 18$ mK 
($ 55 < 1/T < 62$) it is in the CF phase, while for 
$18 < T  < 55$ mK  ($ 20 < 1/T < 55$) it is in the 
EPf phase, as illustrated on Fig.~\ref{fig:PT} for a pure FQH system. 

Here we have to stress that the analysis based on the boundary free
energy is incomplete since we do not account for the contribution 
from the bulk. Nevertheless, the CFT free energy carries information 
about the universal properties of the system, so that it could label 
the FQH universality classes,  and we believe its behaviour
could capture any transition between different FQH phases.

A similar two-step II-nd order phase transition has been proposed for
the $\nu=4/3$ plateau \cite{2-3nay-wilc}. In that case there seems 
to exist an intermediate metallic state  which opens the possibility of a 
two-step transition similar to that described above, i.e., 
a II-nd order transition from the spin-polarized to the intermediate 
partially polarized  compressible phase and
then a II-nd order transition to the spin-polarized state.
This is to be compared with the situation of the  $\nu=2/3$ plateau
where all transitions are I-st order \cite{2-3nay-wilc,mcd-hald}
since a compressible state does 
not exist and because usually II-nd order phase transitions are accompanied
by discontinuity in the compressibility, i.e., any direct transition 
between incompressible states, even if both states have the same 
topological structure, is most likely not of II-nd order. For 
$\nu=2/3$, where the PH-conjugate of the $\nu=1/3$ Laughlin state
and the spin-singlet state seem to have the same topological structure,
it is  confirmed  by numerical calculations \cite{mcd-hald}
that the transition is of I-st order.

The phase transitions described above seem to be ``classical" as
they occur at non-zero temperature. That is probably why 
in the numerical calculations \cite{5-2hr}, performed at $T=0$, 
they only see a second order phase transition MR $\to$ CF 
\footnote{the authors of \cite{5-2hr} cannot conclusively 
determine whether it is  a I-st order or a II-nd order transition.
However, since the transition CF $\to$ EPf is already of first order,
we believe that the discontinuity in the compressibility
rules out the possibility of a I-st order transition MR $\to$ CF} 
and say nothing about any other transition, such as 
CF $\to$ EPf.
Finally, we stress that 
 although we believe that the transition MR $\to$ CF is 
of II-nd order,  there is still a possibility for a smooth crossover 
MR $\to$ CF \cite{5-2hr}, however, the transition CF $\to$ EPf is 
most likely of I-st order.
\section{Abelian versus non-abelian statistics}
The phase transition discussed in the previous section is actually
 a classical transition between abelian and non-abelian FQH states.
It seems that non-abelian statistics is preferred at low temperature,
while at higher temperature the abelian states are more favorable.
Perhaps, this is characteristic for all
FQH plateaux for which abelian and non-abelian states
are competing.
For instance, at $\nu=7/3$ and $8/3$ there exist (PH conjugated)
$k=1$ Laughlin states and $k=4$ parafermion states,
the latter being non-abelian.
Probably, such phase transitions can
explain as well the sharp kinks observed in the activated experiment for
these plateaux \cite{pan}.
Note that, e.g.  for $\nu=7/3$,  the phase transition between the
high-temperature abelian phase (the Laughlin state)
and the low-temperature non-abelian phase (the Read--Rezayi state)
 is of first order, again because of the different topological orders
\cite{mcd-hald}, the absence of an intermediate compressible state
  and the fact that there is no symmetry, which can be spontaneously 
broken.
This is in agreement with  the sharp change of the slope and 
the $y$-intercept of the diagonal resistance in the activation 
experiment \cite{pan}  and is reminiscent of the
first order phase transition between the spin-singlet and
the spin-polarized states at $\nu=2/3$ \cite{mcd-hald,2-3nay-wilc}
which was discussed at the end of the previous section.
\section{Conclusions}
We have described a new universality class relevant for the
FQH state at $\nu=5/2$, determined  by the rational CFT of 
the abelian EPf  state, 
which has a well-defined chiral fermion parity number and 
could be viewed as a $\Z_2$ supersymmetric extension of the MR state.
Using our previous analysis \cite{gaps} we have computed the energy 
gaps, Eqs.~(\ref{gap_EPf}) and (\ref{gap_MR}) for the EPf and MR states,
as well as the periods and amplitudes of the persistent currents
in both states for a disk sample. Based on our analysis
we conclude that there might be a two-step phase transition
between the MR and EPf states at finite temperature,
involving an intermediate compressible state,
in which the chiral fermion parity symmetry is spontaneously
broken. 

In order to reveal the nature of the FQH state at $\nu=5/2$
new and  more precise  experiments are needed. 
The phase transition MR $\to$ EPf investigated in this paper  
could be detected by measuring several quantities for
temperature in the range $10$~mK~$\leq T \leq 30$~mK:
\begin{itemize}
\item{\textbf{Quasiparticle charges:}\\
 The (minimal) quasiparticle charge in both states is different:
 $Q_\mathrm{q.h.}=1/4$ for the MR state and
$Q_\mathrm{q.h.}=1/2$ for the EPf state.
Therefore, a charge-measuring experiment, such as shot-noise, 
for temperatures in the above range could confirm whether the
phase transition seen in the activation experiment \cite{pan}
is the transition between the MR and EPf states described in this paper.
}\\
\item{\textbf{Energy gaps:}\\
The activation energy of the EPf state is significantly bigger than 
that of the MR 
state, i.e.,  $\D_{\EPf}\sim 3 \D_{\MR} $ for the sample of \cite{pan}.
Thus, a more precise activation experiment for a high-mobility sample
like that of \cite{pan} in the above range could confirm our predictions
about the energy gaps and the phase transition. 
\\ }
\item{\textbf{Persistent currents:}\\ 
The ratio of the amplitudes of the persistent currents in the EPf and MR
states at a transition temperature\footnote{we expect that $T_c/T_0$ is 
almost independent of disorder}  $T_c/T_0\sim 1.23$  
is  $I^{\mathrm{EPf}}/I^{\mathrm{MR}}\sim 2.15$.
However, when we take into account the (same) contribution from the 
two $\nu=1$ lowest Landau levels (filled completely with electrons of 
opposite spin), which is estimated to be $2\times 1.07$ in our units 
$ev_F/4L$,  this ratio decreases to $1.06$, which is not detectable with
the current SQUID precision \cite{pers-exp}. Nevertheless, we believe that
when the precision becomes better than $5 \%$, the phase transition 
between the two phases could be practically detected in a high-mobility
sample similar to that of \cite{pan}. 
}
\end{itemize}

In addition, there might be a phase transition at
zero temperature to a BCS-type condensate but
that universality class would be different from both the
MR and the EPf states according to Fig.~\ref{fig:pers},
since the periodicity of the persistent current in
the condensate is expected to be $1/2$ of the flux unit,
while that of the MR and EPf states is always $1$.
Probably, this is described by the strong pairing
phase of \cite{read-green,5-2hr}.

Finally, we believe that all these new experiments, as well as
the analysis in this paper,  may shed more light on the  nature of 
the mysterious FQH state at $\nu=5/2$.
One important aspect of the results presented in this paper
is the anticipation that the non-abelian quasiparticles can be observed 
in practice only for temperatures below the transition temperature,
 which we estimate as $T_c\sim 15$~mK for $\nu=5/2$ in the sample 
of \cite{pan}. 
\begin{ack}
I would like to thank Ivan Todorov, Yakov Shnir,   Kazusumi Ino 
and  especially Michael Geller  for inspiring discussions 
as well as  
the organizers of the NATO Workshop ``Statistical Field Theories",
Como, June 2001 for hospitality and financial support. This work was
supported by DFG  through Schwerpunktprogramm ``Quanten-Hall-Systeme"   
under  the program ``Konforme Feldtheorie der
Quanten-Hall-Plateau-\"Uberg\"ange".
\end{ack}
\begin{appendix}
\section{Weak modular invariance of the EPf state}
\label{app:mod}
In this appendix we are going to show that the characters of the 
EPf state are modular covariant, i.e.,  they belong to a $2$ dimensional 
representation of the  subgroup \cite{cz,fro2000}  
$\Gamma_\theta \subset  PSL(2,\Z)$ of the modular 
group\footnote{this fact is well-known \cite{ginsparg}}. 
We use the explicit form
of the modular $S$-matrix for the Ising model, in the basis of characters
$(\ch_i)=(\ch_0,\ch_{1/16},\ch_{1/2} ) $,
\[
S=\frac{1}{2} \left[ 
\begin{array}{rrr}
1 & \sqrt{2} & 1 \cr 
\sqrt{2} & 0 & -\sqrt{2} \cr
1 & -\sqrt{2} & 1 \end{array}
\right], \quad \mathrm{where} \quad 
\ch_i\left(-\frac{1}{\t}\right)=
\sum_{j=1}^3 \, S_{ij} \, \ch_j(\t)
\]
to show that the Ising factor in the characters (\ref{chi_EPf}) is
$S$-invariant, i.e., 
\beq
\ch_0\left(-\frac{1}{\t}\right)+\ch_{1/2}\left(-\frac{1}{\t}\right)= 
\ch_0(\t)+ \ch_{1/2}(\t).
\eeq 
On the other hand, 
this combination is $T^2$-invariant ($\t\to\t+2$) up to a phase, i.e.,
\[
\left|\ch_0(\t+2)+\ch_{1/2}(\t+2)\right|^2= 
\left| \ch_0(\t) + \ch_{1/2}(\t) \right|^2,
\]
while not being simply  $T$-invariant since 
$|\ch_0(\t+1)+\ch_{1/2}(\t+1)|^2= | \ch_0(\t) - \ch_{1/2}(\t)|^2$.
We recall that the Ising characters are neutral and therefore the
$U:\z\to\z+1$ and $V:\z\to\z+\t$ transformations do not change them.
Thus, the Ising part of the characters is invariant with respect to
$T^2,S,U,V$ transformations and therefore the complete characters 
(\ref{chi_EPf}) of the EPf state have the transformation properties 
of the bosonic $\nu=1/2$ Laughlin state \cite{cz} with respect to
these transformations.
\section{Equivalence of the persistent currents of the EPf state
and the bosonic $\nu=1/2$ Laughlin state}
\label{app:laugh}
According to Eq.~(\ref{pers}) the neutral factor in 
Eq.~(\ref{Z_chi_EPf}), coming from the Ising model, does not contribute
to the persistent current of the EPf state. The latter can be computed
by substituting $\z\to \z+\phi\t$, like in Eq.~(\ref{pers}) and
using the transformation property of the $K$-functions (\ref{K_l})
\beq\label{trans}
\ex^{-\frac{\pi}{m} \frac{\left(\Im(\z +\phi\t)\right)^2}{\Im\t}}
K_l(\t,\z+\phi\t;m)= 
\ex^{-\frac{2\pi i}{m} \left(\frac{\phi^2}{2}\Re\t +\phi \Re\z\right)}
\left(\ex^{-\frac{\pi}{m} \frac{(\Im\z)^2}{\Im\t}} 
K_{l+\phi}(\t,\z;m)\right).
\eeq
In addition one could express the chiral  partition 
function (\ref{Z_chi}) for the $\nu=1/m$ Laughlin state as
\[
Z^+_{\mathrm{Laugh}}(\t,\z)= \ex^{-\frac{\pi}{m} \frac{(\Im\z)^2}{\Im\t} }
\sum_{l \mod m} K_l(\t,\z;m)= 
\ex^{-\frac{\pi}{m} \frac{(\Im\z)^2}{\Im\t} } K_0(\t,\z/m;1/m).
\]
Applying Eq.~(\ref{trans}) to the above equation 
and ignoring the $\z$-independent $\eta$ functions in Eq.~(\ref{K_l}),
we get that the non-zero contribution to the persistent current 
for the $\nu=1/m$ Laughlin state comes 
from\footnote{we set $\Re\t=\Re\z=0$ to guarantee the reality of the 
partition function and choose $\Im\z=0$ since the persistent current is 
periodic in $\phi$}
\beq\label{Z_0}
K_{\phi/m}(\t,0;1/m)\sim \sum_{n\in\Z}q^{\frac{1}{2}\frac{1}{m} (n+\phi)^2}=
\sum_{n\in\Z} \ex^{-\frac{\pi^2}{m} \frac{T_0}{T}(n+\phi)^2},
\eeq
which coincides with Eq. (38) in \cite{geller-loss-kircz}
up to the sign of $\phi$, that  
can be changed by a substitution $n\to -n$. 
Therefore, the persistent current for the 
EPf state is the same as that for the $\nu=1/2$ Laughlin state.
\section{Low-temperature asymptotics of the persistent currents}
\label{app:low-T}
For $T/T_0 \to 0$ the modular parameter vanishes, 
$q=\exp\left(-2\pi^2 T_0/T\right)\to 0$, 
so it is sufficient to keep only the first three terms
$n=0,\pm 1$ in Eq.~(\ref{Z_0}). Thus, for the persistent current 
in the EPf state one gets
\beq\label{IEPF}
I_\EPf(T,\phi)= -\frac{1}{2}\frac{ev_F}{L} \phi+
2\exp\left(-\frac{\pi^2}{2} \frac{T_0}{T}\right)
\sinh\left(\pi^2\frac{T_0}{T} \phi \right), \quad \frac{T}{T_0}<<1, 
\eeq
for $ \vert\phi\vert \leq \frac{1}{2}$. 
The local maximum for $-1/2 \leq \phi \leq 1/2$ at fixed temperature
is located at
\beq\label{phi-max}
\phi^{\max}_{\EPf}(T)=-\frac{1}{\pi^2} \frac{T}{T_0}
\mathrm{arccosh}\left( \frac{1}{2\pi^2} \frac{T}{T_0} 
\exp\left(\frac{\pi^2}{2} \frac{T_0}{T} \right)\right).
\eeq
Substituting Eq.~(\ref{phi-max}) into Eq.~(\ref{IEPF})
and using the identity 
$\sinh(\alpha)=\sgn(\alpha)\sqrt{\cosh^2(\alpha) -1}$ we 
obtain Eq.~(\ref{low-T-EPf}). The same procedure applied to
the MR state gives Eq.~(\ref{low-T-MR}).
\section{High-temperature asymptotics of the persistent currents}
\label{app:high-T}
The high-temperature limit  $T/T_0 \to \infty$ 
is not so trivial because the modular parameter 
$q=\exp\left(-2\pi^2 T_0/T\right)\to 1$ is at the border of the
convergency interval for the partition functions. Therefore it is
more convenient to perform $S$ transformation first
\beq
S:
\begin{array}{l}
\t=-1/ \t' \\
\z=-\z'/ \t' \end{array}\quad 
{\Longleftrightarrow} \quad
\begin{array}{l}
\t'=-1 / \t =i T/\pi T_0 \\ 
\z'=\z /\t =\phi \end{array}.
\eeq
Now the modular parameter 
$q'=\ex^{2\pi i \t'}=\exp\left(-2T/T_0 \right)\to 0$ when
$T/T_0\to \infty$.
Here we use the transformation properties of the characters 
for the Laughlin FQH state with $\nu=1/m$ \cite{cz}
\beq\label{S}
\chi_l(\t,\z)=\exp\left(i\frac{\pi}{m} \Re\frac{{\z'}^2}{\t'}\right)
\sum\limits_{l'=0}^{m-1} S_{ll'} \  \chi_{l'}(\t',\z'),
\quad  S_{ll'}=\frac{1}{\sqrt{m}}\exp\left(-2\pi i\  \frac{ll'}{m}\right).
\eeq
Note that $\Re {\z'}^2/\t'=\phi^2\Re(1/\t')=0$. Next, substitute
Eq.~(\ref{S}) into Eq.~(\ref{Z_chi}) to get
\[
Z^+_\EPf(T,\phi)=\sum_{l'=0}^{m-1}  \chi_{l'}(\t',\z')
\sum_{l=0}^{m-1} S_{ll'} =\sqrt{m} \ \chi_0(\t',\z'),
\]
where we have used that
$\sum_{l=0}^{m-1} S_{ll'}=\sqrt{m} \delta_{l,0} \mod m$.
Now (ignoring $\sqrt{m}$ and  the $\eta$-function) 
keeping only the leading  three terms $n=0,\pm 1$ in 
$\chi_0(\t',\z')$ and taking into account that $\Im\z'=\Im\phi =0$,
i.e., the CZ factors are trivial, we get
\beq\label{log-Z}
\ln Z^+_\EPf(T,\phi) \quad \mathop{\simeq}\limits_{T/T_0 >>1} \quad
\ln\left(1+ 2\cos(2\pi\phi) \exp\left(-m\frac{T}{T_0}\right) \right).
\eeq
The second term is very small in this limit and we use $\ln(1+x)\simeq x$
valid for $x<<1$ to get, after differentiation
with respect to $\phi$ at $\phi^{\max}=-1/4$, Eq.~(\ref{high-T-EPf}) 
for $m=2$. Note that the same result can be obtained from the
$\theta_3$ function formula
 Eq. (39) in Ref. \cite{geller-loss-kircz}
\footnote{the $q$ in Eqs.~(34) and (39) there is $m$
in our notation} .
Indeed, for $T/T_0>>1$ we have 
$\sinh(n m T/T_0)^{-1}\simeq 2\exp(-n mT/T_0)$. Then keeping only the 
first term, $n=1$ in  Eq. (39) in \cite{geller-loss-kircz},
we arrive at the analog  of our  Eq.~(\ref{high-T-EPf}) for general $m$.

We repeat the same calculation for the MR state using the $S$-matrix
computed in \cite{cgt} (see Eq. (5.8) there). The difference is that
now 
\[
\sum_{l=0,\pm 1, \pm 2, 4} S_{ll'}= \left\{ 
\begin{array}{ll}
\sqrt{2}+1, & \quad \mathrm{for}\quad l'=0 \\ 
0, & \quad \mathrm{for}\quad l'=\pm 1, \pm 2 \\
\sqrt{2}-1, &  \quad \mathrm{for}\quad l'=4  \end{array} \right.
\]
so that 
\[
Z^+_\MR(T,\phi) = \left(\chi_0(\t',\z') - \chi_{4}(\t',\z')\right)+
\sqrt{2}\left(\chi_0(\t',\z') + \chi_{4}(\t',\z')\right).
\]
Next we use that for $q'\to 0$ the Ising model characters satisfy
$\ch_0(\t') \pm \ch_{1/2}(\t') \to \ch_0(\t') $ and the identity 
\beqa\label{K_2l}
&&K_{2l}(\t,2\z;8) \pm K_{2l+4}(\t,2\z;8)=
K^{\pm}_{l}(\t,\z;2)= \nn
&&=\frac{1}{\eta(\t)} \sum_{n\in\Z} (\pm 1)^n q^{(n+l/2)^2} 
\ex^{2\pi i \z(n+l/2)}
\eeqa
to get
\[
Z^+_\MR(T,\phi)\quad \mathop{\simeq}\limits_{T/T_0 >>1} \quad
\left(K^-_0(\t',\z';2)+\sqrt{2}K^+_0(\t',\z';2)\right)\ch_0(\t'), 
\]
which after dropping $\z'$ independent factors gives
\[
\ln Z^+_\MR(T,\phi)\quad \mathop{\simeq}\limits_{T/T_0 >>1} \quad
\ln\left( 1 + \frac{\sqrt{2}-1}{\sqrt{2}+1} 2\cos(2\pi\phi) 
\exp\left( -2 \frac{T}{T_0}\right)\right).
\]
Again we use $\ln(1+x)\simeq x$, $x<<1$ and differentiate 
this equation with respect to $\phi$ at $\phi^{\max}=-1/4$, which
gives the high-temperature asymptotics  (\ref{high-T-MR}) of the 
persistent current in the MR state.
\end{appendix}

\def\NP{Nucl. Phys. }
\def\PRL{Phys. Rev. Lett.}
\def\PL{Phys. Lett. }
\def\PR{Phys. Rev. }
\def\CMP{Commun. Math. Phys. }
\def\IJMP{Int. J. Mod. Phys. }
\def\JSP{J. Stat. Phys. }
\def\JP{J. Phys. }
\bibliography{5-2}
\end{document}